\documentclass[%
 aip,
 amsmath,amssymb,
 reprint,%
]{revtex4-1}
\usepackage{graphicx}% Include figure files
\usepackage{dcolumn}% Align table columns on decimal point
\usepackage{bm}% bold math
\usepackage[utf8]{inputenc}
\usepackage[T1]{fontenc}
\usepackage{mathptmx}

\newcommand{\weakpair}[2]{\left\langle  #1 , #2 \right\rangle}
\newcommand{\doubleweakpair}[2]{\left\langle \left\langle  #1 , #2 \right\rangle \right\rangle}

\newcommand{\extcovd}{\extd_{\nabla}}
\newcommand{\dotwedge}{\dot{\wedge}}

\newcommand{\extd}{\textrm{d}}

\newcommand{\volF}{\mu_\text{vol}}

\newcommand{\bound}[1]{i^*({#1})}

\newcommand{\Mbound}{\partial M}

\newcommand{\cl}[1]{\mathcal{#1}}

\newcommand{\tens}{%
  \mathbin{\mathop{\otimes}}%
}

\newcommand{\TwoTwoMat}[4]{
\begin{pmatrix}
#1 & #2 \\
#3 & #4
\end{pmatrix}}
\newcommand{\TwoVec}[2]{
\begin{pmatrix}
#1\\
#2 
\end{pmatrix}}

\newcommand{\pH}{port-Hamiltonian }

\newcommand{\ramy}[1]{{\color{black}#1}}
\newcommand{\DiracEuler}{\mathcal{D}_\text{E}}
\newtheorem{remark}{Remark}
\usepackage{color}

\definecolor{amber}{rgb}{1.0, 0.49, 0.0}

\newcommand{\fps}[1]{{ \color{black}  #1}}

\begin{document}
%\preprint{AIP/123-QED}
\title{Geometric and energy-aware decomposition of the Navier-Stokes equations: A port-Hamiltonian approach}
% Force line breaks with \\
\author{Federico Califano}
\email{f.califano@utwente.nl \\The following article has been submitted to \textit{Physics of Fluids, AIP Journal}. After it is published, it will be found at https://publishing.aip.org/resources/librarians/products/journals.}
\author{Ramy Rashad}%
 
\affiliation{ 
Robotics and Mechatronics Department, University of Twente, The Netherlands%\\This line break forced with \textbackslash\textbackslash
}%

\author{Frederic P. Schuller}
\affiliation{%
Department of Applied Mathematics, University of Twente, The Netherlands%\\This line break forced% with \\
}%

\author{Stefano Stramigioli}
 \affiliation{ 
Robotics and Mechatronics Department, University of Twente, The Netherlands%\\This line break forced with \textbackslash\textbackslash
}%

\begin{abstract}
A port-Hamiltonian model for compressible Newtonian fluid dynamics is presented in entirely coordinate-independent geometric fashion. This is achieved by use of tensor-valued differential forms that allow to describe describe the interconnection of the power preserving structure which  underlies the motion of perfect fluids to a dissipative port which encodes Newtonian constitutive relations of shear and bulk stresses.
The relevant diffusion and the boundary terms characterizing the Navier-Stokes equations on a general Riemannian manifold arise naturally from the proposed construction.
\end{abstract}

\maketitle

\section{Introduction}
Fluid mechanics continues to pose challenging problems of theoretical and practical interest in both the mathematical sciences and engineering applications. This article adds geometric techniques from the theory of open dynamical systems to this toolbox. In particular, it provides a complete geometric decomposition of the Navier-Stokes equations into energy-exchanging subsystems.

The classical Hamiltonian theory of fluid dynamics \cite{Marsden1970,marsden1984semidirect,marsden1984reduction,Morrison1998,Arnold2013}, can only describe conservative systems with no energy-exchange with their environments.
%a fundamental difficulty arises in incorporating the spatial boundary conditions of the system. As a consequence these models tend to focus on conservative systems with no energy-exchange with its surrounding environment.
While this is useful for analysing a system that is isolated from its surroundings, it is an obstacle for any practical applications in which energy exchange with other systems or system components is an inevitable feature, such as the simulation and control of multi-physical systems.
%physical systems.
\fps{A versatile and powerful treatment of such systems requires an extension of Hamiltonian theory, which can deal with open systems and how they are interconnected. The currently best understood such extension is the \pH framework \cite{Geoplex}, which employs a mathematically sophisticated theory of Dirac structures to describe the power exchange between open subsystems.} 
\fps{In} this framework an interaction between two systems is characterised by the 
reciprocal bilateral influence the systems have on each other, and as a consequence
%effect of 
the energy exchanged between them.
This interaction takes place through what is called  a \textit{power port}. Each power port consists of two dual variables, called an effort and flow, whose duality pairing represents the power flowing between the two interacting systems.
The \pH formulation is instrumental to interconnect the fluid dynamic system to other physical systems (of possibly different domains) in a way that intrinsically satisfies the energy continuity between the systems. \fps{This provides a canonical} starting point for the generation of a modular multi-physics network framework 
\fps{that includes Newtonian fluids}.
%able to represent e.g., fluid-solid interconnection systems
%\stefano{as well as advection of electrodynamics properties in the fluid.}

\fps{The straighforward adaptability  of the port-Hamiltonian framework, to which the results of this paper are just further testament, stems from its original purpose to apply general network theory to subsystems provided by dynamical systems. It is now mature in its mathematical foundations and has been applied to a vast variety of dynamical systems. Paramount to our use in this article, } port-Hamiltonian theory has proven very effective in the treatment of distributed-parameter systems \citep{VanDerSchaft2002} (we refer to Ref. \onlinecite{rashad2020twenty} for a recent survey).
%While the traditional Hamiltonian formalism is limited to conservative closed systems, the \pH formalism is applicable to non-conservative open systems capable of energy exchange with its environment. 

The \fps{specific} aim of this work is to extend the port-Hamiltonian formulation of ideal compressible fluids (Euler equations), presented for the first time in Ref. \onlinecite{VanDerSchaft2001}, to Newtonian viscous fluids (Navier-Stokes equations). The way this extension is developed is consistent with the core methodology of \pH modelling: we introduce a distributed port to \textit{interconnect} the system representing ideal fluid dynamics with a purely dissipative system \fps{that encodes} the constitutive relations peculiar of Newtonian fluids.
\fps{This possibility was envisaged} informally in the aforementioned paper\fps{ \cite{VanDerSchaft2001}:} \textit{``Energy dissipation can be incorporated in the framework by terminating some of the ports by a resistive relation. In this way we can represent the Navier Stokes equations.''}  
%We take this challenge seriously, uncovering the geometric structure that is needed to implement this extension in great generality. 
\fps{The present work is a geometrically thorough technical implementation of this vision} in  
\fps{quite useful generality}, namely for \fps{fluid domains represented by arbitrary} Riemannian manifolds of any dimension. \fps{As related work, we cite the much more special case of a flat one-dimensional spatial domain \cite{Altmann2017} where reactive flows were considered}. 
\fps{The way to achieve such a generality is by extending the differential form representation of \pH fluid systems\cite{VanDerSchaft2002} to tensor-valued forms and the identification of novel duality pairings that are necessary to extend the definition of power flows in this context. The employment of tensor-valued forms for the geometric description of stress, in turn, was inspired by the treatment in Ref. \onlinecite{gilbert2019geometric}.} 
While the present paper explicitly only treats Newtonian fluids \fps{in order to keep the presentation focused, it is clear that all relevant constructions lend themselves for extensions} to a larger class of fluids.

\fps{As additional contribution, we use the proposed geometric tools to treat the fluid dynamics together with their thermodynamic evolution, which allows us to draw conclusions on the validity of the proposed \pH model with respect to a general Fourier-Navier-Stokes fluid. }

%The use of differential geometry, and in particular of \textit{exterior calculus}, makes it possible to write equations which are invariant under any change of coordinates. 
\fps{Our consistent use of differential geometric language in the port-Hamiltonian treatment of the Navier-Stokes equations is essential for four reasons} (see Refs \onlinecite{Marsden2012,Arnold2013,Marsden1972} \fps{for an exposition of clodes} fluid dynamical systems in this geometric setting). \fps{First, it ensures that constructions and conclusions are free of involuntary coordinate artefacts, whose absence otherwise had to be proven laboriously on a case by case basis. Secondly, the differential geometric formulation connects to recent results on numerical schemes that exploit algebraic isomorphisms between continuous quantities} and their discretization\cite{Arnold2006} (we refer to e.g. Refs. \onlinecite{Nitschke2017,Mohamed2016} for works in this direction).
\fps{Thirdly,} this provides the right tool to deal with \textit{global} \fps{questions, which cannot be addressed by formulations tied to local coordinate charts. The standard vector calculus representation of the Navier Stokes equations already falls apart if one chooses problem adapted coordinates even in the case of a flat fluid domain. Much more so if} the domain features a non-zero curvature or nontrivial topology.
%euclidean
%\stefano{Euclidean}
%representation of quantities addressed in vector calculus, but represents the only possibility to derive a complete mathematical model of the system in case the topology of the spatial manifold in which the fluid lives is non trivial. 
%It is indeed a fact that vector fields, which roughly represent one of the "state variables" of the fluid dynamic system, do not admit a basis on any smooth manifold. 
Both is already the case for a sphere, a which is the obviously relevant domain for fluid flow in many practical atmospheric models, where coordinate representations can only describe the dynamics on a patch of the domain, \fps{but intrinsically} fails to capture the evolution of the entire vector field as a whole. \fps{Fourthly, and finally,} the \fps{vector calculus} formulation of the Navier-Stokes equations does not contain the information needed for their unique extension to curved fluid domains \cite{Chan2017}, where any \fps{naive} integration of tensor-valued forms is generically meaningless \cite{Marsden1972}.
\fps{In} order to keep the essential methodology clear also for readers which are not practitioners of the framework, we motivate every step \fps{taken towards} the construction of the \pH formulation of the Navier Stokes-equations in this article. 

The paper is organised as follows. In Sec. \ref{sec:coordinateNS} and \ref{sec:geometryNS}  an overview of the Navier-Stokes equations will be presented respectively in the coordinate-dependent and geometric case. In Sec. \ref{sec:energy} the power balance in Newtonian fluids is calculated in the general case, and will help the reader to understand the underlying reasoning behind the definition of a \pH model. The \pH model for ideal fluids is reviewed in Sec. \ref{sec:eulPH} and in Sec. \ref{sec:nsph} the novel model valid for Newtonian fluids is presented. In Sec. \ref{sec:operators} some considerations on the NS equations and the resulting energy balance are drawn. Sec. \ref{sec:conc} contains conclusions and future work. We refer to Appendix \ref{appendixb} for an overview on the complete thermodynamic representation of Newtonian fluids in the proposed differential geometric language.

\paragraph*{Notation.} Standard differential geometric notation is used throughout the article. 
\fps{The symbols used in this paper are as follows.  The mathematical model of fluid domains is a compact, orientable, $n$-dimensional Riemannian manifold $M$ with (possibly empty) boundary $\partial M$. The} space of vector fields on $M$ is the space of sections $\Gamma(TM)$ of the tangent bundle $TM$. The space of differential $p$-forms is denoted \fps{by} $\Omega^p(M)$ \fps{and we also occasionally refer to $0$-forms as functions, $1$-forms as covector fields} and $n$-forms as top-forms. For
$v\in \Gamma(TM)$, we use the standard definitions for the interior product by $\iota_v:\Omega^p(M) \to \Omega^{p-1}(M)$ and the Lie derivative operator $\mathcal{L}_v$ acting on tensor fields of any valence. 
%the exterior derivative $\extd:\Omega^p(M) \to \Omega^{p+1}(M)$ and the Lie derivative $\mathcal{L}_v$ on tensors of any valence. A Riemannian manifold carries a metric $g: \Gamma(TM) \times \Gamma (TM)\to \Omega^0(M)$, which is \stefano{a} positive definite, symmetric and non degenerate $(0,2)$ tensor field. 
The \fps{metric-compatible and torsion-free} covariant derivative $\nabla_v$, the Hodge star operator $\star:\Omega^p(M) \to \Omega^{n-p}(M)$ and the associated volume form $\volF=\star 1$ as well as the musical operators $\flat:\Gamma(TM) \to \Omega^{1}(M)$ and $\sharp:\Omega^{1}(M) \to \Gamma(TM)$, which respectively transform vector fields to 1-forms and vice versa, are all uniquely induced by the Riemannian metric in the standard way. 
When making use of Stokes theorem $\int_M \extd \omega=\int_{\partial M} i^{*}\omega$ for $\omega \in \Omega^{n-1}(M)$, we explicitly employ the linear operator $i^*$, sometimes referred to as the trace operator, which is the pullback of the \fps{canonical} inclusion map $i: \partial M \hookrightarrow M$.
%For example $*_2(\alpha \tens \volF)=\alpha$ (since $\star \volF=\star \star 1 =1$), for any tensor $\alpha$; and $\flat_1(v \tens \beta)=\nu \tens \beta$, where $\nu=\flat(v)=g(v,\cdot)$ and $\beta\in \Omega^p(M)$. 
%We define more advanced operators along the paper when needed. 
When dealing with tensor-valued forms do we adopt the additional convention that a numerical index $i\in\{1,2\}$ on the left or the right of a standard operator on differential forms indicates whether the operator acts on the ``first leg'' (the tensor value, which in this case needs to be a form) or on the ``second leg'' (the underlying form) of the tensor-valued form on the respective side of the operator: For an $n$ form-valued $m$ form $\alpha\otimes\beta$, for instance, we define $\star_1(\alpha \tens \beta):=\star\alpha \tens \beta$ and $\star_2(\alpha \tens \beta):=\alpha \tens \star\beta$. 
For the representation of the above concepts in terms of coordinate charts, see Ref. \onlinecite{gilbert2019geometric}. 
\section{Coordinate-dependent Navier-Stokes equations}\label{sec:coordinateNS}
On flat $n$-dimensional Euclidean space {\it and} after having chosen Cartesian coordinates,
the momentum balance equation for Newtonian fluids can be written as the $n$ equations 
\begin{equation}
\label{eq:momentumBalanceEuclidean}
\partial_t{(\rho v^i)}+\partial_m(\rho v^m v^i)=-\partial^i p  + \partial_m \tau^{mi},
\end{equation}
where all indices range from $1$ to the dimension $n$ of the fluid domain and repeated indices are summed over. On the left hand side of this momentum balance equation, the scalar field $\rho$ represents the mass density of the fluid and the vector field components $v^i$ represent the velocity vector field of the fluid. Both depend on time and their spatial-temporal evolution is related by the supplementary mass continuity equation 
\begin{equation}
\label{eq:ContinuityEuclidean}
\partial_t{\rho}+\partial_m (\rho v^m)=0.
\end{equation}
On the right hand side of the momentum balance equation, the second rank tensor field $\tau$ models the viscous stress of the fluid and is, for Newtonian fluids, equal to
     \begin{equation}        \label{eq:viscousEuclidean}
        \tau^{ij}:=\lambda (\partial_m v^m) \delta^{ij} +\kappa (\partial^i v^j+ \partial^j v^i)
    \end{equation}
in terms of the fluid velocity field and two non-negative constants, the so-called bulk viscosity constant $\lambda$ and the shear viscosity constant $\kappa$. 
% A further specialisation of (\ref{eq:momentumBalanceEuclidean}) to incompressible fluids, i.e. constraining the velocity fields to $\nabla \cdot \bold{v}=0$, produces the famous Laplacian term characterising momentum diffusion in the most celebrated form of Navier-Stokes equations.
Finally, the scalar field $p$ represents the fluid's static pressure and is governed by thermodynamic equations of state corresponding to the specific thermodynamic assumptions one entertains for any given fluid. 
Notice, from the right hand side of the balance equation, that we chose to not encode the pressure $p$ of the fluid as yet another contribution to the stress tensor $\tau$, as it is often done, but instead to keep it as a separate quantity of thermodynamic nature. For definiteness, we consider the extraordinarily simple case of {\it barotropic} fluids, which are compressible fluids whose pressure field $p$ is completely determined by the mass density $\rho$ and the potential $U(\rho)$ for the internal energy density $\rho U(\rho)$ of the fluid, by means of the equation of state
\begin{equation}
\label{eq:barotropicStateEquation}
p=\rho^2 \frac{\partial U}{\partial \rho}(\rho)\,,
\end{equation}
%$$[p] = (M L/T^2)/L^2 = \frac{M}{L T^2}$$
%$$[\rho] = \frac{M}{L^3}$$
%$$[\frac{\partial U}{\rho}] = [U] \frac{L^3}{M}$$
%$$\frac{M}{LT^2} = \frac{M^2 L^3}{L^6 M} [U] $$
%$$[U] = \frac{L^2}{T^2} \qquad\textrm{potential for the internal energy density}$$
%$$[\rho U] = \frac{M}{T^2 L}\qquad\textrm{internal energy per volume / internal energy density}$$
%$$[\int d^3x \rho U] = \frac{M L}{T^2}\qquad\textrm{internal energy}$$
which effectively completely decouples the fluid from the thermodynamic domain. Since the central constructions of the present article are not concerned with thermodynamical issues and, indeed, can be extended to include rather generic thermodynamics assumptions, such as those underlying the Fourier-Navier-Stokes fluid models, the loss of generality incurred by our consideration of barotropic fluids is inessential to the conclusions of this work and 
serves
to keep the essential constructions lean. We will comment on the relation between the proposed model and the complete fluid model including thermodynamics in Appendix \ref{appendixb}.

What buys the above simplicity of formulation of the momentum balance equation (\ref{eq:momentumBalanceEuclidean}), continuity equation (\ref{eq:ContinuityEuclidean}) and the Newtonian viscous stress tensor (\ref{eq:viscousEuclidean}) is only the combination of flatness of the underlying space and the thus enabled choice of cartesian coordinates. This implies particular numerical coincidences: The components of the metric tensor $\delta_{ij}$, of the inverse metric tensor $\delta^{ij}$ and the components $\delta^j{}_i$ of the Kronecker symbol numerically all coincide in these coordinates,
$\delta^{ij} = \delta^i{}_j = \delta_{ij}\,,$
which, in turn, implies that also
$\partial^i := \delta^{ij}\partial_j = \partial_j \delta^{ij} = \partial_i$. 
This simplicity, however, comes at the cost of seriously obscuring the coordinate-independent nature of both the equations and the objects determined by it. The often employed vector calculus formulation of these equations does not repair this at all, since it only hides the indices but does not remove the underlying assumptions of flatness of the underlying space and the need to choose cartesian coordinates on top of that. We will therefore not dwell on the above formulation but replace it in the following section by the already known proper coordinate-independent formulation of the Navier-Stokes equation. Not only does this repair the above-mentioned shortcomings for flat domains but it also directly generalizes the Navier-Stokes equations for fluid flow on a curved domain.
\section{Geometric formulation of the Navier-Stokes equations}\label{sec:geometryNS}

In this section we introduce the coordinate-free formulation of the momentum conservation equation (\ref{eq:momentumBalanceEuclidean}), which can be found e.g., in Refs. \onlinecite{frankel2011geometry,gilbert2019geometric}. 
Whether one adheres to a flat domain for the fluid or generalizes to an $n$-dimensional Riemannian manfold, it does not make a difference for the coordinate-independent formulation that we employ in this work. We  will hence suppose from the beginning that the domain is an $n$-dimensional Riemannian manifold $M$ with metric tensor $g$. Furthermore, in order to \fps{ensure the convergence of all relevant integrals and the applicability of Stokes' theorem, we impose the physically unproblematic assumption} that the manifold is both compact and orientable.
%\footnote{We restrict to a compact manifold because in this way we allow to create larger volumes of fluids by interconnecting compact components, which is the peculiarity of the port-Hamiltonian methodology.} 
The momentum balance equation and the definition of the Newtonian viscous stress tensor as well as the continuity equation take their geometrically most insightful form in this setting when expressed in terms of differential forms of various degrees. Indeed, instead of a time-dependent fluid velocity vector field $v \in \Gamma(TM)$, we rather use the covector field (i.e., $1$-form)
$$\nu := g(v,\cdot)\in \Omega^1(M)$$ 
and instead of the scalar field $\rho$, we employ the mass density top form (i.e., $n$-form)
$$\mu := \star\rho \in \Omega^n(M)$$ %\mu_{\textrm{\tiny vol}}\,,$$
where the operator $\star$ denotes the Hodge dual on the Riemannian manifold $(M,g)$.
%$\mu_{\textrm{\tiny vol}}$ denotes the volume induced by the Riemannian metric $g$. 
The information contained in the viscous stress tensor, finally, is now encoded in a covector-valued $(n-1)$-form $\cl{T}\in \Omega^1(M)\tens \Omega^{n-1}(M)$. 
The motivation of such tensorial nature for stress in this geometric formulation can be extensively found in Refs. \onlinecite{frankel2011geometry,gilbert2019geometric,Toshinwa} and the intuition is that stress, in a continuum, needs to be integrated over a surface to get a traction force, i.e., a covector. We are thus tempted to write the expression for the traction force acting on an $(n-1)$-dimensional surface $S\subset M$ as
   $f_{\cl{T}}=\int_S \cl{T}$
where the integration acts only on the "form part", i.e., the second leg of $\cl{T}$. However, even if we can give a component-wise meaning to this integral on a flat space, integration of tensor--valued forms is not defined on general Riemannian manifolds.
The intuitive reason for this is that the basis vectors on $M$ can change from point to point on a curved space, which does not permit factoring the necessary basis vectors out of the integral. In other terms, the sum/integration of (co)vector belonging to different (co)tangent spaces is not a well defined operation on Riemannian manifolds \cite{Marsden2012,gilbert2019geometric}. 
Instead, the quantity of interest that can be cast into standard integration on manifolds and that will be of fundamental importance in this work is the \textit{rate of work}, or \textit{power}, generated by the stress on a surface. To define it we introduce the useful binary operator:
\begin{equation} \label{eq:DotWedge}
    \dot{\wedge} : (\Omega^1(M) \tens \Omega^{l}(M))\times (\Gamma(TM) \tens \Omega^{k}(M))\to \Omega^{l+k}(M)
\end{equation}
taking as input two tensor valued forms with dual properties on the first leg, which are paired producing a function, while the forms characterising the second leg of the arguments are simply wedged together with the usual $\wedge$ acting on scalar valued differential forms. 
As an example of application of this operator, the following identity which will be used later is valid in case $\alpha\in \Omega^1(M) \tens \Omega^{n}(M)$ and $v\in \Gamma(TM)$:
\begin{equation} \label{eq:DotWedgeIdentity}
 \alpha \dot{\wedge} v = \star \iota_v  \star_2 \alpha,
\end{equation}
where the fluid velocity vector field $v$ is uniquely identified with a vector valued zero-form $v\in \Gamma(TM) \tens \Omega^0(M)$, which is indeed an equivalent way to express a section of the tangent bundle, i.e., a vector field.

We can immediately give a physical interpretation of the power generated by stress on a surface as 
\begin{equation}
    P_{\cl{T}}=\int_S \cl{T}\dot{\wedge} v.
\end{equation} 
Notice that thanks to the covector valued nature of stress, this definition provides a metric independent notion of power.
In order to write the differential momentum equation for a fluid in this language we need a further ingredient to calculate the net force of the stress on a volume element, playing the role of the divergence of the Cartesian version of the stress tensor $\tau$ in (\ref{eq:momentumBalanceEuclidean}). The key operator is the \textrm{exterior covariant derivative} $\extcovd : \Omega^1(M)\tens \Omega^{n-1}(M) \to \Omega^1(M)\tens \Omega^{n}(M)$, combining topological properties of the exterior derivative $\extd$ and metric properties of the Levi-Civita covariant differential $\nabla$ associated to $(M,g)$. Following Refs. \onlinecite{kanso2007geometric,gilbert2019geometric}, an implicit definition of $\extcovd$ as the following identity on the space of top forms $\Omega^n(M)$ is given as
\begin{equation}
\label{eq:defExtcovd}
    (\extcovd \cl{T}) \dotwedge v=\extd(\cl{T}\dotwedge v) - \cl{T} \dotwedge \nabla v,
\end{equation}
for any vector field $v$. Here $\nabla v \in \Gamma(TM) \tens \Omega^1(M)$ is the vector--valued one--form representing the geometric analogous of the \textit{velocity gradient} in Euclidean space, and defined as $\nabla v (X):= \nabla_X v$ for any vector $X$, being $\nabla_X$ the covariant derivative in direction $X$.

In terms of the geometrically well-defined quantities above, the momentum balance equation in convective form on a compact and oriented Riemannian manifold takes the manifestly coordinate-independent form \cite{frankel2011geometry,Marsden2012,gilbert2019geometric}
\begin{equation}
\label{eq:velocityEquationDStress}
      \dot{\nu} +  \extd(\tfrac{1}{2} \star(\nu\wedge\star\nu) )+  \iota_v \extd \nu  = -\frac{\extd p}{\star\mu}+\frac{\star_2 \extcovd \mathcal{T}}{\star\mu}\,.
\end{equation}
The left hand side of (\ref{eq:velocityEquationDStress}) is the differential form representation of the material derivative of the velocity field while the right hand side encodes the way pressure and stress enter as force (covector) fields in the equation.

When considering Newtonian fluids, $\cl{T}$ is composed as the sum of a bulk stress $\cl{T}_{\lambda}$ and a shear stress $\cl{T}_{\kappa}$, defined by
\begin{align}
    \cl{T}_{\lambda} &:= \lambda (\textrm{div}(v)\volF) ,
    \label{eq:DefBulkViscosity}\\
   \cl{T}_{\kappa} &:= 
\kappa (\star_2 \cl{L}_{v}g), \label{eq:DefShearViscosity}
\end{align}
where $\volF$ is the volume form induced by the Riemannian metric. The bulk stress $\cl{T}_{\lambda}$ depends on the divergence of the velocity vector field $\textrm{div}(v)$, defined as the function such that $\cl{L}_{v}\volF=\textrm{div}(v)\volF$ holds true.
The shear stress $\cl{T}_{\kappa}$ is defined in order to model viscous stresses whenever the transport of the metric under the flow of $v$ is non-zero, i.e., when $v$ fails to be the generator of a rigid body motion. In fact $\cl{L}_v g$ extends the concept of \textit{rate of strain}
to Riemannian manifolds. As a matter of fact, in term of components $(\cl{L}_vg)_{ij}=\nabla_i \nu_j +\nabla_j \nu_i$, which is the natural generalisation on manifolds of the rate of strain in (\ref{eq:viscousEuclidean}), constructed by replacing ordinary derivatives with covariant derivatives. It is interesting to notice that, contrarily to the bulk stress, the shear stress does not admit a formulation using scalar valued differential forms. In fact, being $\cl{L}_v g$ a symmetric 2-rank tensor, it cannot be represented by a scalar valued differential form, which is by definition a totally antisymmetric tensor field. This is ultimately the phenomenological reason why a geometric representation of Navier--Stokes equations need to be developed using (co)vector--valued forms.

A computation of the above stress tensor components shows that they encode precisely the same information as $\tau$ in Cartesian coordinates on a flat manifold,
but are now defined on any Riemannian manifold and are manifestly invariant with respect to the choice of coordinates. 

The continuity equation that supersedes (\ref{eq:ContinuityEuclidean}) for the general case of a Riemannian manifold and without coordinate assumptions takes the simple geometric form \cite{VanDerSchaft2002}
\begin{equation}
\label{eq:Continuity}
     \dot{\mu}= - \extd(\star\mu\star\nu),
\end{equation}
where it might aid the intuition to note that the right hand side is identical to minus the Lie derivative $\mathcal{L}_v\mu$. The barotropic equation of state, which we assumed in (\ref{eq:barotropicStateEquation}) for definiteness, is already valid on any Riemannian manifold.

%In this respect the modular approach characterising pH modeling (see e.g. \cite{rashad2020porthamiltonian1}) allows to easily generalise the results in this paper to other fluid models by changing only the thermodynamic potential which generates the pressure. THERE WAS A MORE COMPELLING WAY TO ARGUE WHY IT IS ACTUALLY STRUCTURALLY NOT SO MEANINGFUL TO COUNT THE PRESSURE AS A PART OF STRESS. REMEMBER?
\section{Energy in the fluid and on-shell/off-shell power balance}\label{sec:energy}
The compressible Navier-Stokes momentum balance equation  (\ref{eq:velocityEquationDStress}) and continuity equation (\ref{eq:Continuity}), together with the the barotropic equation of state (\ref{eq:barotropicStateEquation}) we chose for definiteness, describe a system whose energy loss due to viscosity is unavoidable, even if the domain boundaries would not allow for any exchange of matter or energy with the environment. If such an exchange through the domain boundary is additionally possible, the system may experience an additional loss or gain of energy. In any case, the system does not obey conservation of its mechanical energy which is expressed as function of the state variables $\nu$ and $\mu$ as:
\begin{equation}\label{eq:HofNS}
    H(\nu,\mu) := \int_M \left(\tfrac{1}{2}\star\!\mu(\nu\wedge\star\nu) + U(\star\mu)\mu\right)\,.
\end{equation}
composed by the sum of kinetic energy $K(\nu,\mu)=\int_{M} \tfrac{1}{2}\star\!\mu(\nu\wedge\star\nu)$ and the potential energy.
In this section we use the momentum and continuity equations to compute the variation of the total energy in the fluid domain $\dot{H}$, detecting the mechanisms which contribute to a non zero power flow in the system. Such a failure of the energy functional to be a complete generator of the time evolution of a system is the defining hallmark of a system that exchanges energy with its environment. The description of such systems is trivially beyond the scope of generalised Hamiltonian theory and requires a port-Hamiltonian treatment instead. 

Conscious of the fact that port-Hamiltonian theory does not represent yet a consolidated framework in physics and dynamical system theory, we introduce some terminology which aims at helping the reader in understanding the basics of port-based thinking. This terminology is not standard in the port-Hamiltonian literature, but we believe it is extremely useful to understand the role of the key geometric object characterising a pH system, called \textit{Stokes Dirac Structure} (SDS), for non practitioner readers. Consider the energy functional (\ref{eq:HofNS}), dependent on the state variables $\nu$ and $\mu$, also called \textit{energy variables} in the \pH framework. The formal expression for the time derivative of this functional in case of fixed spatial domain (i.e. $M$ does not change in time) is 
\begin{equation}\label{eq:offshell}
   \dot{H}=\int_M \delta_\nu H \wedge \dot{\nu} +\delta_{\mu}H \wedge \dot{\mu},
\end{equation}
where $\delta_{(\cdot)}H$ indicates the variational derivative of $H$ with respect to the energy variables, which correspond to differential forms with a complementary degree with respect to the energy variables and are referred to as \textit{co-energy variables}.
The variational derivatives (see Refs.  \onlinecite{VanDerSchaft2002,rashad2020porthamiltonian1} for a proof) of $H$ with respect to the energy variables are
\begin{equation}\label{eq:varderivs}
	\delta_{\nu}H=\star\mu\star\nu \in \Omega^{n-1}(M) , \delta_{\mu}H=\tfrac{1}{2} \star(\nu\wedge\star\nu) + h \in \Omega^0 (M)\,,
\end{equation}
where $h=\frac{\partial}{\partial \star\mu}(\star\mu\, U(\star\mu))$ is the \textit{specific enthalpy}, which for barotropic fluids is related to pressure by means of the identity 
\begin{equation} \label{eq:enthalpy}
	\extd h=\frac{\extd p}{\star\mu}.
\end{equation}

%In \cite{VanDerSchaft2002} a definition for the variational derivatives in the current framework (i.e., where the energy variables are differential forms) is given. 
Notice that the co-energy variables carry a clear physical meaning, i.e., $\delta_{\nu} H$ is the \textit{mass inflow} over a surface and $\delta_{\mu}H$ is the known as the \textit{Bernoulli function}.
We refer to (\ref{eq:offshell}) as \textit{off-shell} expression of the power, since it does not involve the equations of motion (i.e., the continuity equation and the momentum equation), but only the knowledge on the functional $H$, its energy variables, and the variational derivatives. 
%It basically represents nothing else than the fundamental \textit{energy continuity equation} 
% \begin{equation}
% \dot{\cal H}+d{\iota_n {\cal{H}}}=P \label{eq:ContinuityDiff}
% \end{equation}
% with energy injection $P=0$
% in integral form (Heaviside principle), where ${\cal H}$ represent the top form of energy density, $\iota_n {\cal{H}}$ indicates the flux of energy in direction $n$,  with specific dependency on the fields $\mu$ and $\nu$ using the chain rule. 
In contrast, we call \textit{on-shell} expression of the power, the one obtained by substituting the equations of motion
(\ref{eq:velocityEquationDStress}) and (\ref{eq:Continuity})
in the time derivative of the energy variables in (\ref{eq:offshell}). This distinction, which might seem redundant at first sight, is at the core of pH theory, since the SDS will be defined based on this distinction.
The key to identifying the elements of the port-Hamiltonian description of a system is then to equate the {\it off-shell} expression for the rate of change of the energy functional representing continuity of energy with its {\it on-shell} expression. 
%The former does not use the  equations of motion  of the system, while the latter does. Note by equations of motion, we mean the explicitly stipulated equations of motion for a system, not those obtained from the energy functional, which distinction matters since the two only coincide for systems without energy loss. 

\begin{figure}
	\centering
	\includegraphics[width =0.40\textwidth]{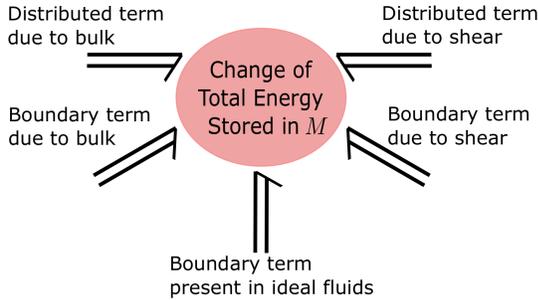}
	\caption{Abstract Energy Balance for Navier-Stokes equations.}
	\label{fig:FNS_Abstract_Balance}
\end{figure}

Now, skipping hereafter the proof which can be found in Appendix \ref{appendix},
equating the off-shell expression (\ref{eq:offshell}) to its on-shell version, i.e., using the equations of motion (\ref{eq:velocityEquationDStress}) and (\ref{eq:Continuity}) in order to replace $\dot\nu$ and $\dot\mu$, we obtain the off-shell/on-shell power balance
\begin{widetext}
\begin{align}
    \underbrace{\int_M \delta_\nu H \wedge \dot{\nu} +\delta_{\mu}H \wedge \dot{\mu}}_{\textrm{Off-shell Term ($\dot{H}$)}} =& \underbrace{\int_{\partial M} i^*(-\star\mu \star \nu) \wedge i^*(\tfrac{1}{2} \star(\nu\wedge\star\nu) +h)}_{\textrm{Boundary Term present also in Ideal Fluids } (\dot{H}_2)}
    +  \underbrace{\int_{\partial M}  i^*(\ramy{\star \cl{T}_\lambda}) \wedge i^*(\star \nu)}_{\textrm{Bulk Boundary term } (\dot{H}_{3\lambda})} + \underbrace{ \int_{\partial M} i^*(\cl{T}_\kappa \dot{\wedge} v)}_{\textrm{Shear Boundary Term } (\dot H_{3\kappa})} 
    +\nonumber \\ -&\underbrace{\lambda\int_M  (\textrm{div}(v))^2 \volF}_{\textrm{Bulk Distributed Term } ( \dot{H}_{3\lambda})} -\underbrace{\kappa \int_M \tfrac{1}{2}  \doubleweakpair{\cl{L}_v g}{\cl{L}_v g}\volF}_{\textrm{Shear Distributed Term } (\dot H_{3\kappa})}
    \label{eq:naiveBalance}
\end{align}
\end{widetext}
where we define the quadratic form on the geometric rate of strain \[\doubleweakpair{\cl{L}_{v}g}{\cl{L}_{v}g}\volF:=(\cl{L}_{v}g)^{\sharp_1}\dot{\wedge}(\star_2 \cl{L}_{v}g).\]
Figure \ref{fig:FNS_Abstract_Balance} summarizes the power balance (\ref{eq:naiveBalance}) for later reference.
The $H_i$ terms in the parenthesis of (\ref{eq:naiveBalance}) refer to the proof of the expression, that can be found in Appendix \ref{appendix}.

As physically expected, both forms of stress produce terms which dissipate mechanical energy within the spatial domain, which are non zero if the divergence of the velocity vector field of the fluid is non zero (bulk) and the rate of deformation of the fluid is non zero (shear).

\begin{remark}
It is important to remark that the shear viscosity also presents a divergence effect in case of compressible fluid, as a consequence of the changing in metric transport, i.e. a vector field $v$ whose divergence is not zero is not the generator for a rigid body motion, and as a consequence $\cl{L}_v g \neq 0$. The combination of both the divergence parts of the bulk and shear components is called \textit{isotropic viscosity}, which has as coefficient the \textit{second viscosity coefficient} $\zeta:= \lambda +\frac{2}{3} \kappa$ (for $n=3$). The remaining part, i.e. the shear viscosity without the divergence term, is called \textit{deviatoric viscosity}. For geometrical consistency we will keep on using the bulk and shear decomposition, since concepts like isotropy are not completely generalisable on Riemannian manifolds.
\end{remark}

A more subtle aspect is that the differential operator encoding stresses in a Newtonian fluid,  also induces terms describing boundary energy flows, which are often neglected in the literature using no-slip boundary arguments ($i^* v=0$) which would indeed nullify the boundary terms in (\ref{eq:naiveBalance}). This argument however is valid only in the case the boundary of the fluid domain $\partial M$ would represent a physical interface between the fluid and a rigid wall on which no slip condition apply. In general we are interested in analysing fluid dynamics also in more abstract situations, e.g., in case the manifold $M$ represents a virtual control volume.

The aim of the \pH approach is to construct the geometric structure, the SDS, able to abstractly capture both the computed energy balance by means of pairings of dual variables, and the equations of motion. In this sense it is a major generalisation of the classical Hamiltonian framework, for which $\dot{H}=0$. The construction will be instrumental for connecting together physical systems in a modular way, keeping the consistency of the power balance induced in the interconnection. 
For this reason it is also important not to discard the boundary terms, because in general, if the fluid is decomposed in different volumes, the no slip condition will not in generally hold on the boundary of such volumes and therefore a description considering every term in (\ref{eq:naiveBalance}) is needed.

In the next section we review the definition of the \pH model for Euler equations, defined in Ref. \onlinecite{VanDerSchaft2002} and derived from Lie group theory in Ref. \onlinecite{rashad2020porthamiltonian1}. Starting from this description we will build the \pH model for NS equations augmenting the reviewed model with a dissipative port encoding the viscous constitutive relation characteristic of Newtonian fluids.
\section{Review of pH formulation of Euler equations with distributed external force field}\label{sec:eulPH}

We review the pH formulation for the barotropic Euler equations, which coincide with the system presented previously by setting the viscous stresses $\cl{T}=0$. The state variables are the velocity 1-form $\nu$ and the mass form $\mu$. The model is generated by the Hamiltonian $H(\nu,\mu)$ in (\ref{eq:HofNS}) which represents the total energy in the fluid dynamic system. The construction builds upon the definition of dual effort and flow variables, which together constitute a power port, in the sense that their duality product represents physical power flowing in the system. The tensorial nature of these variables together with their dual pairing definition is defined in Tab. \ref{tab1} and will be omitted in the text for improving the readability.

\begin{table*}
\begin{tabular}{c|c|c|c|c|c|c|c|c|}
\cline{2-9}
                              & $e_{\nu}$                   & $f_{\nu}$              & $e_{\mu}$               & \textbf{$f_{\mu}$}         & $e_{d}$              & $f_{d}$               & $e_{\partial e}$                    & $f_{\partial e}$                        \\ \hline
\multicolumn{1}{|c|}{Space}   & $\Omega^{n-1}(M)$           & $\Omega^1(M)$          & $\Omega^{0}(M)$         & $\Omega^{n}(M)$            & $\Omega^{1}(M)$      & $\Omega^{n-1}(M)$     & $\Omega^0(\partial M)$              & $\Omega^{n-1}(\partial M)$              \\ \hline
\multicolumn{1}{|c|}{Pairing} & \multicolumn{2}{c|}{$\int_M e_{\nu} \wedge f_{\nu}$} & \multicolumn{2}{c|}{$\int_M e_{\mu} \wedge f_{\mu}$} & \multicolumn{2}{c|}{$\int_M e_d \wedge f_d$} & \multicolumn{2}{c|}{$\int_{\partial M} e_{\partial e} \wedge f_{\partial e}$} \\ \hline
                              & \multicolumn{4}{c|}{Off-shell variables}                                                                    & \multicolumn{4}{c|}{On-shell variables}                                                                                      \\ \cline{2-9} 
\end{tabular}

\caption{\label{tab1} Description of flow and effort spaces and pairings for Euler equations with distributed esternal port $(e_d,f_d)$.}
\end{table*}

The Stokes-Dirac structure $\DiracEuler$ of the fluid dynamic system is given by
\begin{equation}\label{eq:SDS_Euler}
\begin{split}
\DiracEuler = \{ (f_s, &f_{\partial e}, f_d,e_s,e_{\partial e},e_d) \in \cl{B}_\text{E} | \\
			 \TwoVec{f_\nu}{f_\mu} &= \left(\begin{array}{cc}
     \tfrac{1}{\star\mu} \iota_{\sharp\circ\star  (\cdot)} d\nu& d\\
     d & 0  
    \end{array}\right)\TwoVec{e_\nu}{e_\mu}- \TwoVec{\frac{1}{\star\mu}}{0 }e_d,\\
			f_d&= \begin{pmatrix} \frac{1}{\star\mu} &  0\end{pmatrix} \TwoVec{e_\nu}{e_\mu},\\ 
			 \TwoVec{e_{\partial e}}{f_{\partial e}}  &=  \TwoTwoMat{0}{1}{-1}{0}\TwoVec{\bound{e_\nu}}{\bound{e_\mu}}\},
\end{split}
\end{equation}
where the bond space $\cl{B}_\text{E}$ is given by the Cartesian product of the appropriate differential forms which can be uniquely reconstructed using Tab. \ref{tab1}. 
We define
\[f_s=\TwoVec{f_\nu}{f_\mu}\,\,\, \text{and} \,\,\, e_s=\TwoVec{e_\nu}{e_\mu}.\]

\noindent The momentum and continuity dynamic equations
\begin{equation}
\label{eq:EulerEqPort}
\TwoVec{\dot{\nu}}{\dot{\mu}} = \TwoVec{-\extd (\delta_{\mu} H) -  \iota_{v} \extd \nu +e_d/\rho } {-\extd (\delta_{\nu} H)},
\end{equation}
are recovered from (\ref{eq:SDS_Euler}) once we impose 
\begin{align}
\label{eq:SDStoDyn}
e_{s} = \TwoVec{e_{\nu}}{e_{\mu}}=\TwoVec{\delta_{\nu}H}{\delta_{\mu}H}\,\,\, \text{and} \,\,\, f_{s} = \TwoVec{f_{\nu}}{f_{\mu}}=\TwoVec{-\dot{\nu}}{-\dot{\mu}}.
\end{align}
with the variational derivatives in (\ref{eq:varderivs}). Equivalently we say that the implicit port-Hamiltonian system is defined by the inclusion
\begin{equation}
((-\dot{\nu},-\dot{\mu}),f_{\partial e},f_d,(\delta_{\nu}H,\delta_{\mu}H),e_{\partial e},e_d) \in \DiracEuler.
\end{equation}
The assignment (\ref{eq:SDStoDyn}) defines the \textit{off-shell} effort and flow variables, and as such depends only on the Hamiltonian function, the chosen energy variables, and the functional chain rule, which all together express energy continuity as described earlier. In fact (\ref{eq:offshell}) canonically defines dual properties in this effort/flow pair since the variational derivative produces a form of the complementary order of the energy variable along which the variation is calculated (see Tab. \ref{tab1}). The other effort and flow variables in $\DiracEuler$ are the \textit{on-shell} ones, and their definition makes the power balance associated to $\DiracEuler$ consistent with the physical energy balance and produces the correct equations of motion.

The distributed port characterised by the port variables $(e_d$, $f_d)$ is designed such that $e_d$ represents the external force field applied to the momentum balance due to in-domain body forces. Its conjugated variable $f_d$ %=\delta_{\nu}H/\star \mu=\star \nu$ 
represents the volume flux across any surface of the medium and its pairing $\int_M e_d \wedge f_d$ contributes to the time derivative of the Hamiltonian.

The SDS explicits geometrically the power continuity of the system
in the sense that the following balance structurally holds along solutions
\begin{equation}
\label{eq:power_balance}
0=\int_M e_s \wedge f_s+  \int_{\Mbound} e_{\partial e} \wedge f_{\partial e}   + \int_M e_d \wedge f_d.
\end{equation}
Notice that in case Euler equations are considered (i.e., imposing $e_d=0$) this power balance exactly mimics equation (\ref{eq:naiveBalance}) for $\cl{T}=0$.
In the general case (\ref{eq:power_balance}) shows that the variation of total energy in the system is governed by two effects: one pairing of boundary port variables, corresponding to boundary conditions of the underlying PDE, and the pairing of distributed port variables, determining the power flow in the system due to in-domain forces, like body forces (e.g. gravity, magnetic forces) or distributed stress forces. In the following we will address the latter case, showing how to represent the SDS corresponding to Navier-Stokes equations, finally mimicking the whole power balance (\ref{eq:naiveBalance}). 

\begin{remark}
The presented SDS is not standard in the sense of Ref. \onlinecite{VanDerSchaft2002}, due to the state modulated term in the first diagonal entry of the matrix representation of the operator in (\ref{eq:SDS_Euler}). This term is due to convective acceleration of the fluid and a detailed derivation of it can be found in Ref. \onlinecite{rashad2020porthamiltonian1}. Nevertheless the term is skew-symmetric when interpreted as a bilinear operator of the space of off-shell efforts $e_s$, making the term not contribute to the total energy balance of the system. Its effect on the power balance corresponds to the $\dot{H}_1$ term calculated in Appendix \ref{appendix} where (\ref{eq:naiveBalance}) is proven. \end{remark}

\begin{remark}
The possibility of using the ports in the SDS to interconnect the system with other systems is peculiar of the pH framework, whereas in the standard Hamiltonian framework the dynamics is constrained such that no open distributed port is present ($e_d=0$), and the state space is constrained such that the boundary flow, representing mass inflow at the boundary $\partial M$, is zero ($f_{\partial e}=\bound{\iota_{v}\mu}=0$). As consequence the system remains conservative, i.e. $\dot{H}=0$. We will make use of the graphical language of bond graphs (see e.g., Ref. \onlinecite{rashad2020porthamiltonian1} for a complete introduction in this context) to represent dynamical systems interconnected by means of power ports as in Fig \ref{fig:Viscous_Forces_Port}. The \textit{bonds} (double arrows) represent the power ports on which the dual effort and flow variables live and whose pairing defines the power flow in the direction on the arrow. In the standard Hamiltonian framework only the off-shell bond would be present, and as a consequence the power continuity condition encoded by the SDS would collapse to $\dot{H}=0$.
\end{remark}
\section{Construction of pH model for Navier-Stokes equations}\label{sec:nsph}
\newcommand{\effortR}{{\color{black}e_r}}
\newcommand{\flowR}{{\color{black}f_r}}

\begin{figure*}
	\centering
	\includegraphics[width =0.6\textwidth]{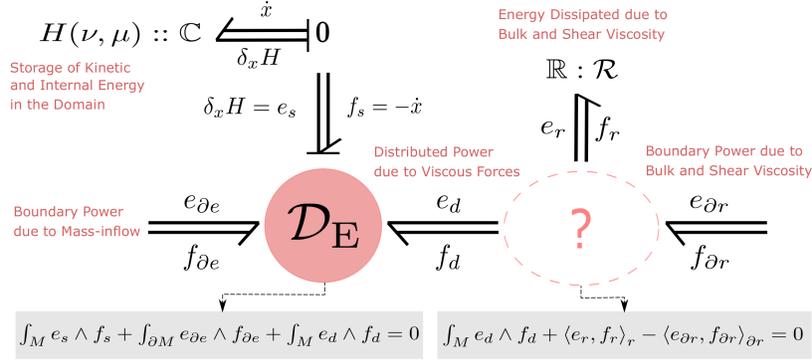}
	\caption{Bond Graph model for the port-Hamiltonian model for Euler's equations with an open distributed port to be used for modeling viscous effects. The dashed arrow shows the power continuity constraint that the SDS encodes.}
	\label{fig:Viscous_Forces_Port}
\end{figure*}

Our aim is to derive the Navier-Stokes equations, together with its geometric SDS, by interconnection. This means that a new resistive port element, with its resistive flow and effort $(e_r,f_r)$ and its duality pairing $\weakpair{e_r}{f_r}_r$ needs to be defined. The viscous constitutive relations in Newtonian fluids must be captured by a resistive static relation $\effortR=\cl{R}(\flowR)$ mapping the resistive effort and the resistive flow such that
\begin{equation}
\label{eq:disspairing}
    \weakpair{\cl{R}(\flowR)}{\flowR}_r\geq 0.
\end{equation}
The sign of the inequality constrains the power flow to be always entering the resistive port, which represents the irreversible transformation of energy to the thermal domain. Therefore, the resistive element cannot generate energy and make it flow back to the rest of the network where it is interconnected.
%\begin{remark}
%This condition, which in case of dissipative systems must clearly hold for physical consistency, is often checked \textit{a posteriori} when one works only at the equation level[cit]. One of the strength of the pH methodology is that this geometrical property is implicitly taken into account and the operators resulting in the equations will automatically satisfy this underlying property.
%\end{remark}
Subsequently the port $(e_d,f_d)$ will be used in order to interconnect the conservative system (Euler equations) to the resistive element in a power preserving way, as depicted in Fig.\ref{fig:Viscous_Forces_Port}. The following interconnection procedure will shed light on the geometric structure underlying the object marked by a question mark, i.e., will determine how the dissipative relation distribute energy in the system.
\subsection{Interconnection procedure}
In case of a closed manifold ($\partial M=\emptyset$), the interconnection will be implemented such that
\begin{equation}
 \int_M e_d \wedge f_d = - \weakpair{e_r}{f_r}_r = - \weakpair{\cl{R}(\flowR)}{\flowR}_r \leq 0,
\end{equation}
meaning that the distributed port $(e_d,f_d)$ is interconnected to the dissipative relation in such a way that the total system will indeed dissipate energy along its solutions. 

The subtle aspect, often treated superficially in the literature with boundary condition arguments, arises in case the compact fluid container has a boundary $\partial M \neq \emptyset$. In this case, a distributed differential operator entering the dynamic equation \textit{induces} a boundary term in the energy balance, which represents the power that might flow in/out the system due to boundary conditions associated to the differential operator. We have seen this mechanism in the calculation of (\ref{eq:naiveBalance}), where the distributed term related to Navier-Stokes equations indeed produces boundary terms into the power balance.  Denoting abstractly these terms by $\weakpair{e_{\partial r}}{f_{\partial r}}_{\partial r}$, the interconnection will be implemented in case $\partial M \neq \emptyset$ as
\begin{align}
\label{eq:interconnectioneulNS}
 \int_M e_d \wedge f_d =& - \weakpair{e_r}{f_r}_r +\weakpair{e_{\partial r}}{f_{\partial r}}_{\partial r}=\nonumber \\ =& - \weakpair{\cl{R}(\flowR)}{\flowR}_r+\weakpair{e_{\partial r}}{f_{\partial r}}_{\partial r},
\end{align}
making the energy balance (\ref{eq:power_balance}) for the total system
\begin{align}
\label{eq:power_balanceBoundary}
\dot{H}=&  \int_{\Mbound} e_{\partial e} \wedge f_{\partial e}   - \weakpair{\cl{R}(\flowR)}{\flowR}_r+\weakpair{e_{\partial r}}{f_{\partial r}}_{\partial r}\leq \nonumber \\\leq& \int_{\Mbound} e_{\partial e} \wedge f_{\partial e}   + \weakpair{e_{\partial r}}{f_{\partial r}}_{\partial r},
\end{align}
which indeed holds true e.g., in (\ref{eq:naiveBalance}) since the distributed dissipative terms are non positive.

As discussed earlier, both the distributed dissipation port $(e_r,f_r)$ and its corresponding boundary port $(e_{\partial r} ,f_{\partial r})$ comprise of two stress contributions due to bulk and shear stresses.
We define $\weakpair{e_r}{f_r}_r:=\weakpair{e_{\lambda}}{f_{\lambda}}_{\lambda}+\weakpair{e_{\kappa}}{f_{\kappa}}_{\kappa}$ and $\weakpair{e_{\partial r}}{f_{\partial r}}_{\partial r}:=\weakpair{e_{\partial\lambda}}{f_{\partial\lambda}}_{\partial\lambda}+\weakpair{e_{\partial\kappa}}{f_{\partial\kappa}}_{\partial\kappa}$.
Thus, we can re-express (\ref{eq:interconnectioneulNS}) as
\begin{align}
	 \int_M e_d & \wedge f_d 
	 = - \weakpair{e_r}{f_r}_r +\weakpair{e_{\partial r}}{f_{\partial r}}_{\partial r} = \nonumber \\
	 & \underbrace{- \weakpair{e_\lambda}{f_\lambda}_\lambda +\weakpair{e_{\partial \lambda}}{f_{\partial \lambda}}_{\partial \lambda}}_{\textrm{Bulk viscosity}}
	 \underbrace{- \weakpair{e_\kappa}{f_\kappa}_\kappa +\weakpair{e_{\partial \kappa}}{f_{\partial \kappa}}_{\partial \kappa}}_{\textrm{Shear viscosity}}.\label{eq:Decomposed_Viscosity_Balance}
\end{align}

In what follows, we will guide the reader in deriving the port-variables and their corresponding pairings in (\ref{eq:Decomposed_Viscosity_Balance}).
We will analyse the two constitutive equations for bulk and shear viscosity separately.
In this way it will be easier to identify the variables playing roles in the energy balance which are generated by the specific constitutive relation under study, in favour of a modular construction of the desired model, in which a user can just "pick up" the desired constitutive relation.

Another advantage of this splitting is that the bulk viscosity relation can be represented using purely scalar valued differential forms, making its port description easier.
On the other hand, defining the resistive relation for the shear viscosity will be more challenging and are non-trivially related to the distributed port $(e_d,f_d)$.
In particular, the shear resistive port variables will be represented by covector-valued forms and thus lie in different spaces and will be coupled with a different pairing with respect to the usual wedge pairing characterising the $(e_d,f_d)$ port. This represents a substantial generalisation with respect to the seminal work \cite{VanDerSchaft2002} where only scalar valued forms are considered.

%We analyse the two constitutive equations separately. In this way it will be easier to identify the variables playing roles in the energy balance which are generated by the specific constitutive relation under study, in favour of a modular construction of the desired model, in which a user can just "pick up" the desired constitutive relation. Another advantage of this splitting is that the bulk viscosity relation can be represented using purely scalar valued differential forms, making its port description easier.
For both stresses, the strategy we use to define their corresponding port variables and pairings relies on the power balance of the system (\ref{eq:naiveBalance}), following from the phenomenological way in which Newtonian stress forces enter the momentum equation on a continuum.

\subsection{Bulk Viscosity}
\newcommand{\effortLam}{{\color{black}e_{\lambda}}}
\newcommand{\flowLam}{{\color{black}f_{\lambda}}}

\begin{figure*}
	\centering
	\includegraphics[width =0.8\textwidth]{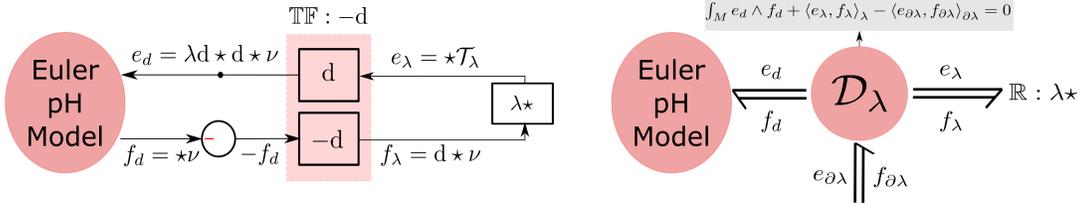}
	\caption{Augmenting the Euler pH model with bulk viscous forces. Left figure shows block diagram representation in case $\partial M = \emptyset$, while Right figure shows bond graph representation in the general case.}
	\label{fig:Bulk}
\end{figure*}

The comparison between the bulk terms in (\ref{eq:naiveBalance}) and the generic interconnection to be achieved (\ref{eq:interconnectioneulNS}) instructs us on how to define the resistive effort and flow variables generated by the bulk dissipative relation. These are of both distributed type, denoted  $e_{\lambda}$ and $f_{\lambda}$, and of boundary type, denoted $e_{ \partial \lambda}$ and $f_{ \partial \lambda}$. The static resistive relation is denoted $\mathcal{R}_{\lambda}$. The construction is achieved by defining the on-shell effort and flow variables $e_{\lambda},f_{\lambda},e_{\partial \lambda},f_{\partial \lambda}$, their pairings, and $\cl{R}_{\lambda}$ such that
\begin{align}
   -\weakpair{e_{\lambda}}{f_{\lambda}}_{\lambda} &=-\lambda\int_M  (\textrm{div}(v))^2 \volF \label{eq:bulkmimic1} \\
   \weakpair{e_{\partial \lambda}}{f_{\partial \lambda}}_{\partial \lambda} &= \int_{\partial M}  i^*(\ramy{\star \cl{T}_\lambda}) \wedge i^*(\star \nu).
   \label{eq:bulkmimic2}
\end{align}

\begin{table}
	\centering
\begin{tabular}{c|c|c|c|c|}
\cline{2-5}
                              & $e_{\lambda}$                  & $f_{\lambda}$               & $e_{\partial \lambda}$                          & $f_{\partial \lambda}$                           \\ \hline
\multicolumn{1}{|c|}{Space}   & \ramy{$\Omega^{0}(M)$}                & \ramy{$\Omega^n(M)$}               & $\Omega^{0}(\partial M)$                        & $\Omega^{n-1}(\partial M)$                       \\ \hline
\multicolumn{1}{|c|}{Pairing} & \multicolumn{2}{c|}{$\int_M e_{\lambda} \wedge f_{\lambda}$} & \multicolumn{2}{c|}{\textbf{$\int_{\partial M} e_{\partial \lambda} \wedge f_{\partial \lambda}$}} \\ \hline
\end{tabular}
\caption{\label{tabbulk} On-shell bulk viscosity flow and effort spaces and pairings}
\end{table}

% The following construction is based on defining the aforementioned new variables in order to make the power balance
% \begin{equation}
% \label{eq:bulkpowerbalance}
% \dot{H}= - \weakpair{e_{\lambda}}{f_{\lambda}}_{\lambda}+\weakpair{e_{\partial \lambda}}{f_{\partial \lambda}}_{\partial \lambda}
% \end{equation}
% \textit{mimic} the power balance (\ref{eq:bulkpowerbalance}). It will then be possible to define a geometric object, the SDS, which governs the energy balance of the system in its full generality.

The definition of the proper spaces and pairing achieving the goal is shown in Tab. \ref{tabbulk}.
Furthermore we have $\cl{R}_\lambda:\Omega^n(M) \to \Omega^0(M) $ such that $\effortLam=\cl{R}_\lambda(\flowLam)=\lambda \star \flowLam$, which makes the inequality (\ref{eq:disspairing}) hold true for $\lambda\geq 0$.
For what concerns the on-shell variable assignments we define $\flowLam=\extd \star \nu$, $e_{\partial \lambda}=\bound{\ramy{\star \cl{T}_\lambda}}$ and $f_{\partial \lambda}=\bound{\star \nu}$. It is easily verified that this choice satisfies (\ref{eq:bulkmimic1},\ref{eq:bulkmimic2}).

In Figure \ref{fig:Bulk} the details on the construction of the power preserving interconnection between the ideal fluid port $(e_d,f_d)$ and the bulk dissipative port $(e_{\lambda},f_{\lambda})$ is depicted: the distributed flow \ramy{$-f_d=-\star \nu$} is transformed in $\flowLam$ by means of the operator \ramy{$-\extd$}, and enters the dissipative constitutive relation $\cl{R}_{\lambda}$ to generate $\effortLam=\ramy{\lambda \star \extd \star \nu = \star \cl{T}_\lambda}$ (which corresponds exactly to the bulk stress tensor). This variable comes back to the system by means of the operator $\extd$, producing $e_d=\lambda \extd \star \extd \star \nu$, which is exactly, up to division by $\star \mu$, what pops up at the level of momentum equation due to the bulk viscosity. 

The fact that (\ref{eq:Decomposed_Viscosity_Balance}) holds for this construction implies that \ramy{the operator $\extd$ is \textit{formally skew adjoint} (or just \textit{skew adjoint}} in case $\partial M=\emptyset$) with respect to the defined duality pairings. \ramy{With reference to Fig. \ref{fig:Bulk}, the maps $-\extd$ and $\extd$ form together what is called in bond-graphs and network theory a \textit{Transformer} $\mathbb{TF}$, which is a power continuous element relating efforts and flows.
The Dirac structure $\cl{D}_\lambda$, shown in Fig. \ref{fig:Bulk} (right), implements all the flow-effort relations described above such that the power balance is as desired in (\ref{eq:Decomposed_Viscosity_Balance}), excluding the shear viscosity.
}The boundary variables, if interpreted as PDE boundary conditions, do not enter the momentum equation but have an effect on the energy balance.

%\begin{figure}
%    \centering
%\includegraphics[scale=0.6]{Figures/Bulk.pdf}
%    \caption{Caption}
%    \label{fig:Bulk}
%\end{figure}

\begin{remark}
Notice that this scheme resembles exactly the \textit{grad/div} relation often used to describe constitutive equations together with conservation laws, and the term present at the momentum equation $\extd \star \extd \star \nu$ is indeed the manifold generalisation of the vector calculus term $\nabla (\nabla \cdot \mathbf{v})$ present in compressible Navier-Stokes equations on Euclidean space.
\end{remark}

\subsection{Shear Viscosity}
\newcommand{\effortKap}{{\color{black}e_{\kappa}}}
\newcommand{\flowKap}{{\color{black}f_{\kappa}}}

We proceed analogously as for the bulk viscosity case. The comparison between the shear terms in (\ref{eq:naiveBalance}) and (\ref{eq:interconnectioneulNS}) impose the constraints

\begin{align}
   -\weakpair{e_{\kappa}}{f_{\kappa}}_{\kappa} &=-\kappa \int_M \tfrac{1}{2}  \doubleweakpair{\cl{L}_vg}{\cl{L}_vg}\volF \label{eq:shearmimic1} \\
   \weakpair{e_{\partial \kappa}}{f_{\partial \kappa}}_{\partial \kappa} &= \int_{\partial M} i^*(\cl{T}_\kappa \dot{\wedge} v).
   \label{eq:shearmimic2}
\end{align}

\begin{figure*}
	\centering
	\includegraphics[width =0.8\textwidth]{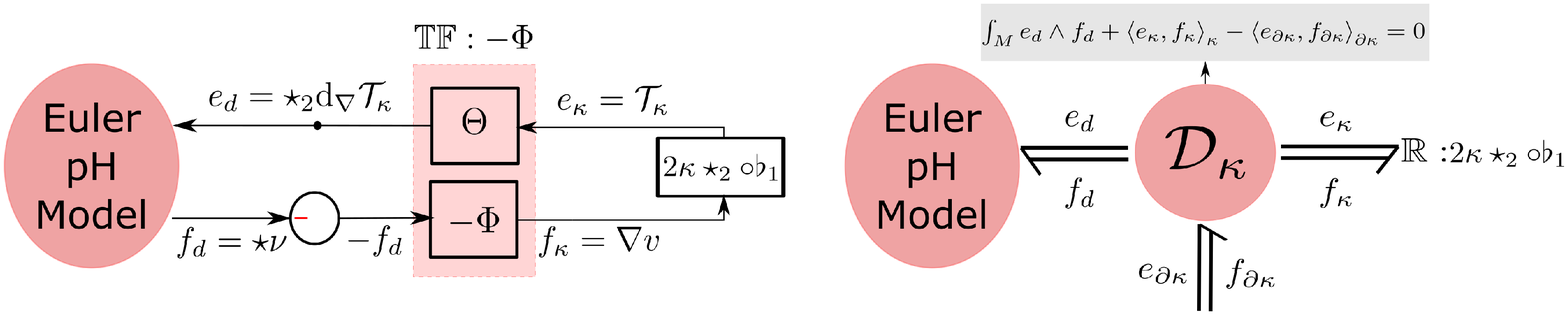}
	\caption{Augmenting the Euler pH model with shear viscous forces. Left figure shows block diagram representation in case $\partial M = \emptyset$, while Right figure shows bond graph representation in the general case.}
	\label{fig:Shear}
\end{figure*}

\begin{table*}
	\centering
	\begin{tabular}{c|c|c|c|c|}
		\cline{2-5}
		& $e_{\kappa}$                      & $f_{\kappa}$                   & $e_{\partial \kappa}$                                        & $f_{\partial \kappa}$          \\ \hline
		\multicolumn{1}{|c|}{Space}   & \ramy{$\Omega^1(M) \tens \Omega^{n-1}(\partial M)$} & \ramy{$\Gamma(TM) \tens \Omega^1(M)$} & $\Omega^1(M) \tens \Omega^{n-1}(\partial M)$         & $\Gamma(TM) \tens \Omega^0(\partial M)$       \\ \hline
		\multicolumn{1}{|c|}{Pairing} & \multicolumn{2}{c|}{$\int_M e_{\kappa} \dot{\wedge} f_{\kappa}$}   & \multicolumn{2}{c|}{$\int_{\partial M} e_{\partial \kappa} \dot{\wedge} f_{\partial \kappa}$} \\ \hline
	\end{tabular}
	\caption{\label{tabshear} On-shell shear viscosity flow and effort spaces and pairings}
\end{table*}

on the definition of on-shell shear variables. The notation goes exactly like in the bulk case with a subscript $\kappa$ instead of $\lambda$. As anticipated, this time the distributed and boundary dissipative ports must be characterised by pairings which go beyond the framework of scalar valued forms. We propose a construction in which the on-shell shear variables and their pairings are shown in Tab. \ref{tabshear}.
We define $\cl{R}_\kappa:\Gamma(TM) \tens \Omega^1(M) \to \Omega^1(M) \tens \Omega^{n-1}(M)$ such that $\effortKap=\cl{R}_\kappa(\flowKap)=2 \kappa \star_2 \circ \flat_1 (\flowKap)$. This choice makes the inequality (\ref{eq:disspairing}) hold true for $\kappa\geq 0$. The distributed on-shell flow is defined as $\flowKap=\nabla v$.

For what concerns the boundary port variables we cannot proceed as in the bulk viscosity case since the pull back does not trivially distribute over $\dot{\wedge}$, but only over the form part, that is the second leg, of its arguments, i.e., $i^*(\cl{T}_{\kappa} \dot{\wedge} v)=i^*_2 (\cl{T}_{\kappa}) \dot{\wedge} i^*_2 (v)$. Here $i^*_2$ denotes the pullback on the second leg of its argument and produces a \textit{two point tensor}, in particular $i^*_2(\cl{T}_{\kappa}) \in \Omega^1(M) \tens \Omega^{n-1}(\partial M)$ and $i^*_2(v) \in \Gamma(TM) \tens \Omega^{0}(\partial M)$. The boundary on-shell variable assignment is then defined as
$e_{\partial \kappa}=i^*_2(\cl{T}_{\kappa})$ and $f_{\partial \kappa}=i^*_2(v)$ and the pairing in Tab. \ref{tabshear} is well defined since the definition of $\dot{\wedge}$ can be canonically extended for two point tensors.
As for the bulk case, this choice satisfies (\ref{eq:shearmimic1},\ref{eq:shearmimic2}).

We complete the picture referring to Figure \ref{fig:Shear}, in which the details about the implementation of the interconnection are given. The distributed flow \ramy{$-f_d=-\star \nu$} is transformed in $\flowKap$ by means of \ramy{minus} the operator $\Phi: \Omega^{n-1}(M)\to \Gamma(TM) \tens \Omega^1(M)$, defined as $\Phi(\star \nu)=\nabla v$, and enters the dissipative constitutive relation $\cl{R}_{\kappa}$ to generate $\effortKap=\kappa \star_2 \mathcal{L}_{v}g$ (which is exactly the shear stress tensor $\cl{T}_{\kappa}$). This variable comes back to the system by means of the operator $\Theta:=\star_2 \extcovd$, producing the right term (up to division by $\star \mu$) appearing in the momentum equation $e_d=\star_2 \extcovd \cl{T}_{\kappa}$. In the same sense as in the bulk case, the operators $\Theta$ and $-\Phi$ can be considered formally adjoint with respect to the defined duality pairings.
\ramy{The combination of $\Theta$ and $-\Phi$ represent a transformer element, but now with a new pairing on one side.}

\begin{remark}
The reason why the defined boundary variables are actually two point tensors whose first leg carries information on $M$, and not $\partial M$, is deeply encoded in the nature of shear stress. In fact, in order to compute the power flow at the boundary in (\ref{eq:shearmimic2}), one needs to know the velocity field in a whole open neighbourhood of the boundary, and then pull back the power density produced by the pairing of stress and velocity field. From a practical point of view, the pairing can be performed using limiting arguments, i.e. calculating a combination of spatial derivatives of the velocity field at the boundary on some coordinate chart. However, geometrically, these derivative, which are those encoded in the rate of strain tensor $\cl{L}_v g$, do not possess a purely topological expression in differential forms that can be pulled back at the boundary, which makes the use of two point tensor necessary to achieve technical correctness.
\end{remark}

\begin{remark}
Notice that theses geometric constructions presents some degrees of freedom in the interplay between the definition of the pairings and of the resistive efforts and flows. The choice we made is to define the quantities such that every variable has a clear physical interpretation, e.g., the distributed resistive efforts represent the stress tensors.
\end{remark}

%\begin{figure}
%    \centering
%\includegraphics[scale=0.6]{Figures/Shear.pdf}
%    \caption{Caption}
%    \label{fig:Shear}
%\end{figure}

\subsection{Port-Hamiltonian model of Navier-Stokes equations}
\begin{figure}
	\centering
	\includegraphics[width =0.475\textwidth]{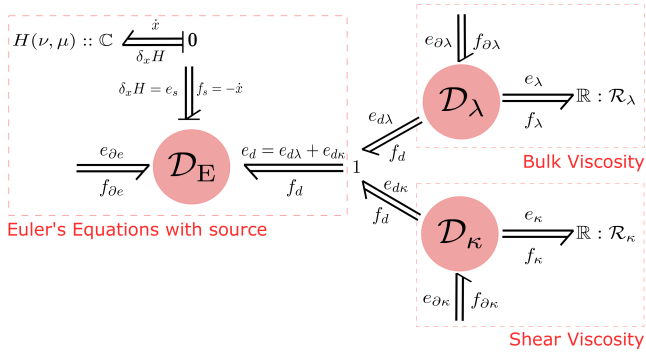}
	\caption{Full decomposed bond graph model of the Navier-Stokes equations on a general Riemannian manifold with boundary.}
	\label{fig:Comp_Isen_viscosity_pH_Bond_Graph}
\end{figure}

\begin{figure}
	\centering
	\includegraphics[width =0.44\textwidth]{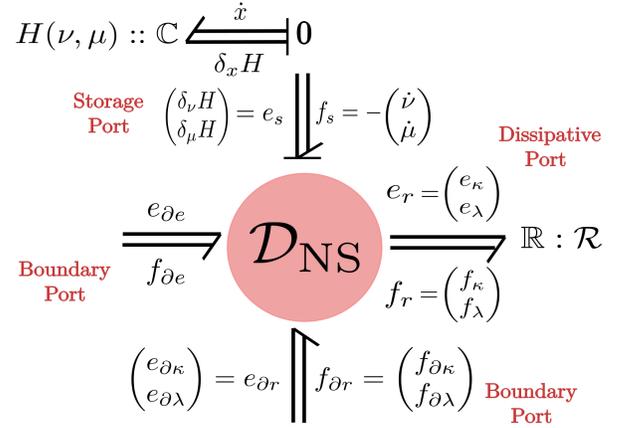}
	\caption{Compact bond graph model of the Navier-Stokes equations on a general Riemannian manifold with boundary.}
	\label{fig:Comp_Isen_viscosity_pH_Bond_Graph_Compact}
\end{figure}

We are now fully prepared to define the overall port-Hamiltonian model for viscous Newtonian fluids, represented using bond graphs as an interconnection of components in Fig. \ref{fig:Comp_Isen_viscosity_pH_Bond_Graph} and compactly in Fig. \ref{fig:Comp_Isen_viscosity_pH_Bond_Graph_Compact}.

To summarise the construction, the following SDS $\cl{D}_\text{NS}$ (in contrast with the previous $\DiracEuler$) constitutes the new geometric entity completely characterising Navier-Stokes equations and the power balance of a Newtonian fluid on a Riemannian manifold. We indicate with
\[f_r:=\TwoVec{f_{\kappa}}{f_{\lambda}}, e_r:=\TwoVec{e_{\kappa}}{e_{\lambda}}, f_{\partial r}:=\TwoVec{f_{\partial\kappa}}{f_{\partial\lambda}}, e_{\partial r}:=\TwoVec{e_{\partial\kappa}}{e_{\partial\lambda}}\]

\[\cl{R}:=\textrm{diag}\{\cl{R}_\kappa,\cl{R}_\lambda\}, \Theta_r:=(\Theta \,\,\, \, \ramy{\extd}), \Phi_r:=\TwoVec{\Phi}{\ramy{\extd}}\]

\noindent and $\cl{B}_\text{NS}$ is the appropriate bond space, which can be uniquely derived by means of the tables. The new SDS \ramy{that combines $\DiracEuler, \cl{D}_\lambda,$ and $\cl{D}_\kappa$} is given by
\begin{equation}\label{eq:SDS_NS}
\begin{split}
\cl{D}_\text{NS}= \{ (f_s, f_r, &f_{\partial e}, f_{\partial r}, e_s, e_r, e_{\partial e}, e_{\partial r}) \in \cl{B}_\text{NS} | \\
			 \TwoVec{f_\nu}{f_\mu} &= \left(\begin{array}{cc}
     \tfrac{1}{\star\mu} \iota_{\sharp\circ\star  (\cdot)} d\nu& d\\
     d & 0  
    \end{array}\right)\TwoVec{e_\nu}{e_\mu}- \TwoVec{\frac{1}{\star\mu} \circ \Theta_r}{0 }\effortR,\\
			\flowR&= \begin{pmatrix} \Phi_r \circ \frac{1}{\star\mu} &  0\end{pmatrix} \TwoVec{e_\nu}{e_\mu}, \\
			\TwoVec{e_{\partial e}}{f_{\partial e}}  &=  \TwoTwoMat{0}{1}{-1}{0}\TwoVec{\bound{e_\nu}}{\bound{e_\mu}},\\
			  \TwoVec{e_{\partial \lambda}}{f_{\partial \lambda}}  &=  \TwoVec{\bound{\star f_\lambda}}{\bound{ e_{\nu}}},\\
			  \TwoVec{e_{\partial \kappa}}{f_{\partial \kappa}}  &=  \TwoVec{i^*_2(f_\kappa)}{i^*_2(\sharp \circ \star (e_{\nu}))}
			\
			\}
\end{split}
\end{equation}
where the resistive static relation $\effortR=\cl{R}(\flowR)$ is imposed.
The structure is power continuous, in the sense that the following power balance folds true along solutions
\begin{align}
\label{eq:power_balanceTot}
\dot{H}=&  \int_{\Mbound} e_{\partial e} \wedge f_{\partial e} + \weakpair{e_{\partial r}}{f_{\partial r}}_{\partial r}  - \weakpair{e_r}{f_r}_{r}\nonumber \leq \\ \leq& \int_{\Mbound} e_{\partial e} \wedge f_{\partial e} + \weakpair{e_{\partial r}}{f_{\partial r}}_{\partial r},
\end{align}
which mimics exactly (\ref{eq:naiveBalance}) once the effort and flow definitions are substituted.
As simple and physical consistent corollary we have that in case of closed manfifold ($\partial M=\emptyset$) the system is purely dissipative, i.e. $\dot{H}\leq 0$, and the dissipation stops when both the rate of deformation $\cl{L}_{v}g$ and the divergence $\textrm{div}(v)$ of the fluid are zero.
\section{Operators, Vorticity and Power Balance in Navier-Stokes equations}\label{sec:operators}
As custom in port-Hamiltonian theory, the explicit dynamic equations can be derived after \textit{eliminating} from the model the explicit dissipative elements. This is done by directly imposing the dissipative static relation in the dynamic equations after imposing (\ref{eq:SDStoDyn}).
The total dynamic equations result in 
\begin{equation}
\label{eq:NStotalEqu}
\TwoVec{\dot{\nu}}{\dot{\mu}} = \TwoVec{-\extd (\delta_{\mu} H) -  \iota_{v} \extd \nu +\frac{\Theta_r \circ \cl{R} \circ \Phi_r}{\rho^2}(\delta_{\nu} H) } {-\extd (\delta_{\nu} H)}.
\end{equation}
While the continuity equation remains rightfully unchanged with respect to Euler equations, it is interesting to look at the momentum equation in its full generality.
By substituting the variational derivatives and the definitions of the maps, and in particular using the form $\delta_{\nu}H=\rho \star \nu$, we obtain the momentum equation which is based on operators acting on the velocity field:
\begin{equation}
    \dot{\nu} +\extd(\tfrac{1}{2} \iota_v \nu)+\iota_v\extd \nu+\frac{\extd p}{\rho} =\frac{\lambda}{\rho} \extd \star \extd \star \nu+\frac{\kappa}{\rho}\star_2 \extcovd \star_2 \mathcal{L}_vg
\end{equation}
The right hand side in the previous equation stores in great generality all possible extra terms that the NS equations might assume with respect to Euler equations on a Riemannian manifold. 
\subsection{The Laplacian operator}
It is instructive to reduce the equation to the incompressible case, in order to have a feeling on the comparison with respect to the
%laplacian
Laplacian
operator, ubiquitous in Navier-Stokes equations. The \pH formulation of ideal incompressible fluids \cite{rashad2020porthamiltonian2} is slightly different with respect to the model presented for compressible fluids due to the different role of pressure. Here the considerations done on the equations are unimpressed by this difference, and hence we safely use the incompressible case for discussing some relevant aspects. 
It is insightful to recall the intuitive geometric idea of the Laplacian operator which is an operator calculating the difference of a field value at a point from ``an average of the field'' on a ``shell of points around it''. Clearly, in a non-flat space, the definition of this ``shell'' can be done in different ways, which explains intuitively the fact that there are different choices of Laplacian that can be made at a manifold level (we refer to the insightful discussion in Ref. \onlinecite{Chan2017} and references therein).

In the incompressible case $\textrm{div}(v)=\star \extd \star \nu=0$. The relation between the proposed geometrical formulation and the 
%laplacian
Laplacian
operator is disentangled by the identity (see Ref. \onlinecite{gilbert2019geometric} for a proof)
\begin{equation}
\label{eq:RicciLaplacian}
\star_2 \extcovd \star_2 \mathcal{L}_v g=\Delta_R\nu
\end{equation}
where $\Delta_R\nu:=\Delta \nu +2\textrm{Ric}(v)$ is the so called \textit{Ricci laplacian}, being $\Delta=-(\star \extd \star \extd + \extd \star \extd \star)$ the \textit{Hodge laplacian} acting on differential forms, and $\textrm{Ric}: \Gamma(TM) \to \Omega^1(M)$ is the \textit{Ricci tensor}, a $(0,2)$ tensor field storing curvature information of $M$ and completely characterised by the metric $g$.
Thus, the incompressible and homogeneous (setting $\rho=1$) NS equations on a Riemannian manifold can be written as
\begin{equation}
\label{eq:incNS}
    \dot{\nu} +\extd(\tfrac{1}{2} \iota_v \nu)+\iota_v\extd \nu+\extd p=\kappa(\Delta \nu +2\textrm{Ric}(v)),
\end{equation}
which assume the familiar form used in vector calculus (identifying the hodge laplacian with the vector calculus laplacian acting on vector components) once a flat Euclidean space as underlying manifold $M$ is considered, i.e. when $\textrm{Ric}=0$. 
In this case, the ``averaging'' around the point can be seen to be equal to an averaging on a surface of dimension $n-1$ with radius tending to zero.
This result is consistent with many accepted versions of incompressible NS equation on manifolds derived on the basis of operator theory only\cite{Chan2017,Kobayashi2008}.

As further insight, we rewrite the momentum equation using (\ref{eq:RicciLaplacian}) as
\begin{align}
    \dot{\nu} &+\extd(\tfrac{1}{2} \iota_v \nu+h+\frac{(\kappa-\lambda)}{\rho}\star \extd \star \nu)= \nonumber \\  =&-\iota_v\extd \nu-\frac{\kappa}{\rho} \star\extd \star \extd \nu+\frac{2 \kappa}{\rho} \textrm{Ric}(v).
\end{align}
One of the advantages of the velocity representation of fluid dynamic systems, is that the vorticity equation, i.e. the dynamic equation governing the closed vorticity $2$-form $\omega:=\extd \nu$ and playing the role of the curl of the velocity field in vector calculus, is simply obtained by exterior differentiating both sides of the velocity equation. The result is
\begin{equation}
    \dot{\omega}=-\mathcal{L}_v\omega-\nu\extd(\frac{\star\extd \star \omega}{\rho})+ \frac{2 \extd \textrm{Ric}(v)}{\rho}
\end{equation}
which in the incompressible and homogeneous case becomes 
\begin{equation}
    \dot{\omega}=-\mathcal{L}_v\omega+\nu \Delta \omega+  2\extd \textrm{Ric}(v).
\end{equation}
As a consequence, only on a flat space, the incompressible vorticity equation reduces to a pure advection-diffusion equation: 
\begin{equation}
    \underbrace{\dot{\omega}+\mathcal{L}_v\omega}_{\textrm{advection}}
    =
    \underbrace{\nu \Delta \omega}_{\textrm{diffusion}}.
\end{equation}
\subsection{The role of vector-valued forms in the energy balance}
In the following we give an energy based argument on why the developed geometric SDS is important to represent Newtonian fluid dynamics by showing that using only the equations can lead to a non physically consistent interpretation of the energy balance. Let us restrict to the incompressible and homogeneous case for simplicity, even if the whole argument is equally valid for compressible fluids.
Using (\ref{eq:incNS}) (which follows from (\ref{eq:RicciLaplacian})), we can compute the power flow due to shear stresses in an alternative way with respect to how it was calculated in Appendix \ref{appendix}, as (the notation is consistent with the proof of (\ref{eq:naiveBalance}))
\[\dot H_{3\kappa}=\kappa \int_{ M} \star \nu \wedge (\Delta \nu + 2\textrm{Ric}(v)).
\]
Using the incompressibility condition and \textit{formal self-adjointness} of the hodge laplacian operator we get the expression
\[\dot H_{3\kappa}=-\kappa \int_{ M} [\omega \wedge \star \omega +\star \nu \wedge 2\textrm{Ric}(v)] + \kappa \int_{\partial M} i^*( \nu \wedge \star \omega),
\]
where we remind that $\omega=\extd \nu$ is the vorticity 2-form. 
In this form it possible to see a dissipative term that depends explicitly on the curvature of the underlying manifold, 
%how 
as
experimentally observed in e.g., Ref. \onlinecite{Debus2017}. Equating the two versions of the power balance we obtain the interesting identity
\begin{align*}
&-\kappa \int_M \tfrac{1}{2}  \doubleweakpair{\cl{L}_v g}{\cl{L}_v g}\volF+\int_{\partial M} i^*(\cl{T}_\kappa \dot{\wedge} v)=\\ &=-\kappa\int_{ M} [\omega \wedge \star \omega +\star \nu \wedge 2\textrm{Ric}(v)] +  \kappa \int_{\partial M}  i^*(\nu \wedge \star \omega).
\end{align*}
We produce the following considerations:
\begin{enumerate}
    \item Using this second version of power balance it could in principle be possible to represent the shear dissipative ports using scalar valued forms only, as clearly seen from the right side of the identity. However this would come to the price of loosing the geometric nature of the stress tensor in the \pH formulation. Furthermore the distributed term $\int_M \star \nu \wedge \textrm{Ric}(v)$ is indefinite in sign, depending whether the curvature of the manifold is positive or negative, which makes this extra distributed port non necessarily dissipative. The reason is that this port has no independent physical meaning at an energy balance level, and makes sense only together with the other terms. We conclude that the definition of such a port would not correspond to a physical mechanism underlying a constitutive relation, and indeed be against the concept of modular interconnection of subsystems.
    \item The identity reveals an interesting connection between the energy balance, the underlying topological properties of the spatial manifold, and the boundary terms. Consider for example a subset of an Euclidean space as $M$, i.e., $\textrm{Ric}=0$, and a fluid undergoing a rotational rigid body motion, i.e., the vector field $v$ satisfies $\cl{L}_v g=0$ and has a constant in space non zero vorticity $\omega$.  Of course the left hand side of the identity must be zero because both the distributed and the boundary term are zero. Nevertheless the vorticity $\omega$ is not zero for a rotational rigid body motion and since $\textrm{Ric}=0$ it must be that \[ \int_{ M} \omega \wedge \star \omega  =   \int_{\partial M}  i^*(\nu \wedge \star \omega),\] which means that the boundary port in this alternative representation acts as a correcting term to make the energy balance right, at the price, again, of loosing physical insight on the stress. This example proves that even in flat space the terms on the right hand side of the identity do not represent physical distributed and boundary dissipation, but just two terms whose sum gives the right result. Again, building the ports in the SDS on the basis of this alternative energy balance based on operators in the NS equation would be wrong, since the ports would not represent the way of physically interconnecting the system to other systems.
    \item In case of closed manifold $\partial M =\emptyset$ we can deduce the kinematic identity 
    \[\tfrac{1}{2}  \int_M \doubleweakpair{\cl{L}_v g}{\cl{L}_v g}\volF=\int_M [\omega \wedge \star \omega +\star \nu \wedge 2\textrm{Ric}(v)],\]
    where we notice that the fact that manifold is closed implies that $\textrm{Ric}\neq0$.
\end{enumerate}
As a conclusion, it is of utmost importance to start with the correct geometric definition of stress in order to define the SDS in a physical consistent way, and these arguments confirm that identifying the model with the equations only can lead to misinterpretation of the terms in the energy balance.
\section{Conclusions and future work}\label{sec:conc}
In this paper, we obtained a coordinate-invariant, port-Hamiltonian formulation of the Navier-Stokes equations for compressible Newtonian fluids. In particular, we showed how the geometric modelling of shear stresses, in terms of  tensor-valued forms and exterior covariant derivatives, allows to describe a novel geometrical duality within the port-Hamiltonian formalism that allows to describe the constitutive relations of Newtonian fluids.
This result opens up the possibility to interconnect Newtonian fluid models with any %interconnection with 
other physical system in order to realise an entirely modular and multi-physical network. Most importantly, this modelling methodology that is based on port-Hamiltonian system theory, ensures that the fundamental physical requirement of overall energy conservation is adhered under all circumstances.

A generalisation of the here obtained model, which includes moving fluid domains in order to be able to interconnect fluid patches to solid mechanics, is under study. The aim in reach is a complete energy-consistent fluid/solid system that can be understood in terms of interconnected open boundary ports of the two subsystems. 
Furthermore we are working towards a \pH description of the complete Fourier-Navier-Stokes fluid, described with the proposed differential geometric language in Appendix \ref{appendixb}, where we discuss the thermodynamic ranges in which the presented model is valid. 
%he upcoming extension is preliminary addressed in terms of representation of the internal energy with a two-port storage element, in which entropy constitutes a state variable. The resistive element  will be substituted with an $RS$ element, which represents in the pH framework the irreversible transformation of energy to the thermal domain.
Finally, a practical application of the proposed framework is the development of advanced numerical integration techniques based on the thorough exterior calculus formulation and the intrinsic preservation of the energy-consistency of the continuous model. The promise is for this to lead to a versatile energy consistent and naturally multi-physical simulation framework.

%Finally, even if the treatment of the Navier-Stokes equations has been done for non thermally isolated fluids only, any kind of viscous effect will generate entropy and therefore the barotropic assumption is a simplification going against basic thermodynamic principles. A formulation has been then shown on how it is possible to generalise the given description by recovering standard thermodynamics extensive variables including entropy with its advection, and generation of it due to temperature gradients and irreversible transformation of mechanical energy from viscous effects. T

%\renewcommand\thesection{A}

\appendix
\section{Proof of (\ref{eq:naiveBalance})}
\label{appendix}
Let us decomposes the on-shell rate of change of energy into a sum  $\dot{H}_1 + \dot{H}_2 + \dot{H}_3$ of conveniently chosen terms, which we now discuss in turn. 
Using the identity (\ref{eq:enthalpy}), we identify the on-shell power expression by substituting (\ref{eq:velocityEquationDStress}) and (\ref{eq:Continuity}) in (\ref{eq:offshell}):
\begin{align}
\dot{H}&=\int_M
     \delta_\nu H \wedge[
     \underbrace{-  \extd (\tfrac{1}{2}\star(\nu\wedge\star\nu))}_{H_2}
     \underbrace{-  \iota_v \extd \nu}_{H_1}  
     \underbrace{-\extd h}_{H_2}
     \underbrace{+\frac{\star_2 \extcovd \mathcal{T}}{\star\mu})]}_{H_3} 
     + \nonumber \\ &+
     \delta_{\mu}H \wedge 
     \underbrace{- \extd(\star\mu\star\nu)}_{H_2}
     \label{eq:terms}.
\end{align}
The underbrace notation identifies the single power terms that are analysed separately. 
Using the identity $\iota_{v}\extd \nu=\star(\star \extd \nu \wedge \nu)$, we take as the first term $\dot{H}_1:=- \int_M \delta_{\nu}H \wedge \star(\star \extd \nu\wedge \nu)$ and show that it vanishes. Indeed, using (\ref{eq:varderivs}) and that $\star\alpha\wedge\beta = (-1)^n \alpha\wedge\star \beta$ for arbitrary $k$-forms $\alpha,\beta$, one obtains $\dot H_1 = (-1)^n\int_M (\star\mu) \nu \wedge \star \extd \nu\wedge\nu = 0$. 

As second term we consider $\dot{H}_2:= - \int_M [\delta_{\nu}H \wedge  \extd(\star(\tfrac{1}{2} \nu\wedge\star\nu)+h)+\delta_{\mu}H \wedge \extd(\star\mu \star\nu)]$. Substituting the variational derivatives (\ref{eq:varderivs}), suing the distribution of the exterior derivative on the wedge product, and using Stokes' theorem, one obtains a mere surface term
%get $\dot{H}_1=\int_M (- \rho * \nu) \wedge \extd((1/2 \iota_v \nu)+h)-\extd((1/2 \iota_v \nu)+h) \wedge (\rho * \nu)$. Using the Leibniz rule "backwards" for the exterior derivative and Stokes theorem it yields \[
\begin{equation}
\dot{H}_2=\int_{\partial M} i^*(-\star\mu \star \nu) \wedge i^*(\tfrac{1}{2} \star(\nu\wedge\star\nu) +h)\,,
\end{equation}
where the pullback $i^*$ of the canonical inclusion map $i: \partial M \hookrightarrow M$ was distributed over the wedge product.

The third term collects the remaining contribution $\dot{H}_3:=\int_M \delta_\nu H \wedge \frac{\star_2 \extcovd \mathcal{T}}{\star \mu}$ to the power balance. Note that this is the term whose presence is synonymous with dissipation due to viscosity and thus effects all the difference to a Eulerian fluid, in which energy variation can indeed happen only at the boundary of the spatial domain. We address the bulk 
(\ref{eq:DefBulkViscosity})
and shear (\ref{eq:DefShearViscosity})
stress separately by splitting the term $H_3$ in $H_{3\lambda}+H_{3\kappa}$. 

%because they are of different nature: the shear stress does include a shear divergence effect, but its effect on the stress will be complementary to the bulk viscosity divergence term which will use a different viscous coefficient $\lambda$. 
For any smooth function $f$ we have $\extcovd(f \volF)=\extd f \tens \volF$, which follows from the Leibniz rule together with the fact $\extcovd \volF=0$. As a consequence we have
\[\star_2 \extcovd \cl{T}_{\lambda}=\lambda \star_2(\extd (\textrm{div}(v))\tens \volF)=
\lambda \extd \star \extd \star \nu,\]
where we used the identity $\textrm{div}(v)=\star \extd \star \nu$. This allows to compute the energy balance for the bulk viscosity term using only scalar valued differential forms as 
\[ \dot{H}_{3\lambda}=\lambda \int_M \star \nu \wedge \extd \star \extd \star \nu. \] 
An integration by parts together with Stokes theorem produces
\[
\dot{H}_{3\lambda}=
\lambda \int_{\partial M} i^*(\star \nu) \wedge 
i^*( \star \extd \star \nu)
-
\lambda \int_M \underbrace{\extd \star \nu}_{\textrm{div}(v)  \volF}
\wedge 
\underbrace{ \star \extd \star \nu}_{\textrm{div}(v)},
\]
which are the bulk terms in (\ref{eq:naiveBalance}) since $\lambda \textrm{div}(v)=\star \cl{T}_{\lambda}$.

For the shear stress we substitute the variational derivative and apply the identity
$\star \iota_v \alpha= ( \nu \wedge \star \alpha)$ for $\alpha \in \Omega^1(M)$ after $\star \nu \wedge \alpha= \nu \wedge \star \alpha$ in the integral, where in this case $\alpha=\star_2 \extcovd \cl{T}_{\kappa}$. Then we use the definition of the operator $\dot{\wedge}$ introduced in
(\ref{eq:DotWedge}) with the identity (\ref{eq:DotWedgeIdentity}). Overall one obtains
\[\dot{H}_{3\kappa}=\int_M \star \nu \wedge \star_2 \extcovd \cl{T}_{\kappa}= \int_M \star \iota_v  \star_2 \extcovd \cl{T}_{\kappa}= \int_M \extcovd \cl{T}_{\kappa} \dot{\wedge} v,
\]
and thus, applying the definition (\ref{eq:defExtcovd}) and Stokes theorem
\[\dot H_{3\kappa}=\int_{\partial M} i^*(\cl{T}_\kappa \dot{\wedge} v) -\int_M  \cl{T}_\kappa \dot{\wedge} \nabla v.\]
The shear terms in (\ref{eq:naiveBalance}) are recovered by the identity $\nabla v=\tfrac{1}{2}(\cl{L}_{v}g)^{\sharp_1}$ (see \cite{gilbert2019geometric} for a proof), 
and the definition $\doubleweakpair{\cl{L}_{v}g}{\cl{L}_{v}g}\volF:=(\cl{L}_{v}g)^{\sharp_1}\dot{\wedge}(\star_2 \cl{L}_{v}g)$.

\renewcommand\thesection{B}
\section{The role of thermodynamics: towards a Fourier-Navier-Stokes port-based model}
\label{appendixb}
In this section we present an overview on the entropy production in fluid flows using the proposed differential geometric framework and then comment on the relation with the proposed pH model. The goal is to highlight rigorously the physical and mathematical steps that are normally, more or less clearly, assumed to include thermodynamic aspects in the Navier-Stokes framework, leading to the so called model for Fourier-Navier-Stokes fluids. 

For a thermodynamic system, the first law assumes the existence of a total energy $H_{t}=\int_M \cl{H}_t$, $\cl{H}_t\in \Omega^n(M)$ which is conserved at an integral balance level, that is

\begin{align}
\label{eq:energyConservation}
    \dot{H}_t&=-\int_{\partial M} i^*(\iota_v \cl{H}_t)  -\int_{\partial M} i^*(\iota_v(\star p)) +\nonumber  \\ &+ \int_{\partial M} i^*( \cl{T}\dot{\wedge} v)-\int_{\partial M} i^*Q
\end{align}
where all the terms on the right hand side correspond to advection of total energy through the boundary, i.e., according to the first law, the total energy does not have sources or sink inside the spatial domain $M$. The first term on the right hand side corresponds to the total energy advected through the boundary by means of a macroscopic velocity field $v$,
the second term expresses the external work done by static pressure, the third term to work done by stress $\cl{T}$, and the fourth the heat flux vector $Q \in \Omega^{n-1}(M)$.

Applying Stokes theorem and definition (\ref{eq:defExtcovd}) we get the energy equation

\begin{equation}
\label{eq:energyEquation}
    \dot{\cl{H}}_t=-\extd(\iota_v \cl{H}_t +i_v (\star p)+Q) + \extcovd \cl{T} \dot{\wedge} v+\cl{T} \dot{\wedge} \nabla v,
\end{equation}

which completes the dynamic equations governing the fluid together with momentum and continuity equations. Newtonian stresses are already characterised by the previously discussed geometric relations for Newtonian fluids while the missing constitutive relation is the one for the heat flux

\begin{equation}
    Q=k \star \extd T,
\end{equation}
where $T \in \Omega^0(M)$ is the temperature field of the fluid and $k$ is the heat conductivity coefficient. Notice that the total energy $H_t$ is \textit{not} the Hamiltonian used before to construct the port-Hamiltonian theory, which comprised only the recoverable energy (i.e., the \textit{Gibbs free energy}). This can be clearly seen by noting that, for simplicity considering a manifold without boundary, from (\ref{eq:energyConservation}) it follows $\dot{H}_t=0$ (total energy is conserved anyway, also if stresses are present), while in the port-Hamiltonian model we had $\dot{H}\leq0$. 
The total energy $H_t$ is the sum of a kinetic energy $K=\int_{M} \cl{K}$, where $\cl{K} =\tfrac{1}{2}\star\!\mu(\nu\wedge\star\nu) \in \Omega^n(M)$ (which is the same used in the Hamiltonian $H$) and an internal energy $U_i=\int_{M} \cl{U}(V,S)$ which depends on the extensive variables of \textit{massic volume} $V:=\star(\frac{1}{\rho}) \in \Omega^3$ and on the extensive thermodynamic state variable \textit{entropy} $S\in \Omega^3(M)$.
In a state of thermodynamic equilibrium it is assumed that the so called \textit{Gibbs relation} is valid, relating the  entropy and other thermodynamic potentials
\begin{equation}
\label{eq:Gibbs}
T \delta S=\delta \mathcal{U}+ p \delta V.
\end{equation}

The "differential" $\delta$ cannot be substituted with exterior derivative operator, since it would produce a collapsing identity on $n+1$ forms. The assumption which needs to be done to proceed in this context is to \textit{assume} that (\ref{eq:Gibbs}) holds also for non-equilibrium conditions in which a macroscopic velocity field $v$ is present. This is implemented by postulating the version of Gibbs equation
\begin{equation}
\label{eq:GibbsConv}
T \frac{DS}{Dt} =\frac{D\cl{U}}{Dt} + p \frac{D V}{Dt} 
\end{equation}
in which the convective derivative operator is $\frac{D}{Dt}(\cdot):=\frac{\partial}{\partial t} (\cdot)+\cl{L}_v (\cdot)$, which takes this form in the differential geometric context due to the geometric version of Reynolds transport theorem \cite{frankel2011geometry}.
Now the dynamic equation on the internal energy $\cl{U}$ is obtained by \textit{subtracting} from (\ref{eq:energyEquation}) the time derivative of the kinetic part, i.e., $\dot{\cl{U}}=\dot{\cl{H}}_t-\dot{\cl{K}}$. The latter is easily computed by means of the momentum and continuity equations and it results in 
\begin{equation}
    \dot{\cl{K}}=-\extd (\iota_v\cl{K}+\iota_v(\star p))+(\extd \star \nu)p+ \extcovd \cl{T}\dot{\wedge} v.
\end{equation}
After computing the subtraction we get the equation on the internal energy
\begin{equation}
\label{eq:internalEnergy}
    \dot{\cl{U}}=-\extd (\iota_v\cl{U}+Q)-(\extd \star \nu)p+\cl{T} \dot{\wedge} \nabla v ,
\end{equation}
rightfully indicating that the distributed stress power $ \cl{T} \dot{\wedge} \nabla v$ is a source for the internal energy of the fluid.
Now applying Cartan's formula together with the fact that $\cl{U}$ is a top form, (\ref{eq:internalEnergy}) can be rewritten as
\begin{equation}
   \frac{D \cl{U}}{Dt} =-\extd Q-(\extd \star \nu)p+\cl{T} \dot{\wedge} \nabla v.
\end{equation}
Now we use (\ref{eq:GibbsConv}) as a definition for the entropy. In fact substitution of previous expression therein, together with the fact (following from the continuity equation)
$\frac{D}{Dt}V=\extd \star \nu$, yields
\begin{equation}
\label{eq:entropy}
   T \frac{DS}{Dt}=-\extd Q + \cl{T} \dot{\wedge} \nabla v.
\end{equation}
To derive a very insightful form of this equation let's apply the identity $\frac{\extd Q}{T}=\extd(\frac{Q}{T})+\frac{Q}{T^2} \wedge \extd T$ and massage (\ref{eq:entropy}) in
\begin{equation}
    \dot{S}+\extd(\iota_v S +\frac{Q}{T})=\sigma_{T}+\sigma_{\cl{T}}
\end{equation}
where $\sigma_T=\frac{k}{T^2} \star \extd T \wedge \extd T\geq 0$ is the entropy source due to a non null temperature gradient while $\sigma_{\cl{T}}=\frac{1}{T} \cl{T} \dot{\wedge} \nabla v \geq 0$ is the entropy source due to non conservative stresses. Notice that the positiveness of the entropy source terms follow in this setting by the choice of Newtonian constitutive relations for the stress and $k>0$, but in case of more complicated fluids it represents a \textit{constraint} for the constitutive relations that can be used, since a negativeness of those terms would constitute a violation of the second principle.

The proposed pH model does not of course capture the whole Fourier-Navier-Stokes picture discussed in the previous section. In fact the thermodynamic equation (\ref{eq:internalEnergy}) as separate conservation law is not part of the model, and the total power balance encoded in the SDS characterises only the variation of the Gibbs free energy (in contrast to the total energy $H_t$) under specific thermodynamic constraints. In the proposed model the constraint is implicit in the assumption of barotropic fluids, for which the potential mechanical energy density $U(\rho)$ (in contrast to the total internal energy density $\cl{U}$) completely characterises the pressure by means of the equality $\extd h=\frac{\extd p}{\rho}$. This assumption represents an instance of the Gibbs relation in equilibrium condition and differs from the more general (\ref{eq:Gibbs}). The research on the geometric port-based structure of the complete Fourier-Navier-Stokes fluids is being intensively investigated (see e.g., Ref. \onlinecite{vanderschaftThermo} and references therein) and it is a known fact that it does not possess a canonical pH formulation, due to the non symplectic nature of the thermal domain. In particular, as evident from the entropy equation, the presence of dissipative stress acts as source of entropy in the fluid, making the assumption of barotropicity false for a thermodynamically isolated system. As a consequence, the proposed model is valid for a non isolated system, in which a low source of entropy keeps the entropy inside the fluid container constant. From an engineering point of view, this scenario comprises all the situations of open low speed aerodynamics systems, in which the pressure of the system can be uniquely derived from the potential energy; or most commonly in the incompressible case, in which the pressure does not have a thermodynamic nature and the NS equations are effectively decoupled from the thermodynamic domain. We remind that a pH formulation of the latter case without viscosity (incompressible Euler equations) is addressed in Ref. \onlinecite{rashad2020porthamiltonian2} and the upgrade to the incompressible \pH Navier-Stokes formulation follows exactly like in this work by considering the shear stress $\cl{T}_{\kappa}$ only.

% \begin{figure}
%     \centering
% \includegraphics[scale=0.6]{Figures/FNS-BG.pdf}
%     \caption{Complete Energetical decomposition of the Fourier-Navier-Stokes Equations using the Bond-Graphs notation in which the half arrows indicate an energy flow, positive or negative, represented by the introduced duality products.}
%     \label{fig:FNS-BG}
% \end{figure}

% \todo{A sketch of the complete energetic model for Fourier Navier Stokes fluids is depicted with the bond graph language in Fig.\ref{fig:FNS-BG}, where, differently to the previously built pH model, the dissipative element is replaced by an RS element, a dual port element which represents the irreversible transformation of energy to the thermal domain. Furthermore also the internal energy storage element is a dual port element since both entropy and massic volume represent the energy variable for the functional.
% We stress that in case general interconnection of fluid patches need to be implemented (such as merging control volumes), all these boundary terms need to be taken into consideration in order to obtain a physically consistent interconnection in terms of energy continuity. A complete characterisation in terms of SDS of this model will be treated in a future work.}

\section*{Availability of data}
Data sharing is not applicable to this article as no new data were created or analysed in this study.

\section*{Funding}
This work was supported by the PortWings project funded by the European Research Council [Grant Agreement No. 787675]

%\nocite{*}
\bibliography{refs}% Produces the bibliography via BibTeX.

\end{document}